\documentclass[a4paper,reqno]{amsart}
\usepackage[T1]{fontenc}
\usepackage[english]{babel} 

\usepackage{amsmath,amssymb,amsfonts,amsthm}
\usepackage{mathrsfs, esint, dsfont}
\usepackage{moreverb,rotating,graphics,float}
\usepackage{caption, subcaption}
\usepackage{tikz}
\usepackage{graphicx}
\usepackage{pgfplots}
\pgfplotsset{compat=newest}
\usetikzlibrary{shapes, calc}
\usepackage[colorlinks, linkcolor={blue!50!blue}, citecolor={red}]{hyperref}
\usepackage{framed,fancybox}
\usepackage{enumerate}
\usepackage{csquotes}
\usepackage{todonotes}

\newtheorem{theorem}{Theorem}
\newtheorem{proposition}[theorem]{Proposition}
\newtheorem{remark}[theorem]{Remark}
\newtheorem{lemma}[theorem]{Lemma}
\newtheorem{corollary}[theorem]{Corollary}

\newtheorem{conjecture}[theorem]{Conjecture}

\newcommand\1{{\mathds 1}}

\def\C{{\mathbb C}}

\def\bbI{{\mathbb I}}

\def\R{{\mathbb R}}

\def\SS{{\mathbb S}}

\def\bk{{k}}

\def\bx{{x}}

\def\bone{{\mathbf 1}}

\def\bnull{{\mathbf 0}}

\def\rd{{\mathrm{d}}}
\def\re{{\mathrm{e}}}
\def\ri{{\mathrm{i}}}

\def\cE{{\mathcal E}}
\def\cF{{\mathcal F}}
\def\cG{{\mathcal G}}
\def\cH{{\mathcal H}}

\def\cK{{\mathcal K}}
\def\cL{{\mathcal L}}

\def\cN{{\mathcal N}}

\def\cV{{\mathcal V}}

\newcommand{\ii}{\infty}
\newcommand{\dps}{\displaystyle}
\newcommand{\eps}{\varepsilon}
\newcommand{\nn}{\nonumber}

\newcommand{\tr}{{\rm tr}}
\newcommand{\spinup}{\uparrow}
\newcommand{\spindown}{\downarrow}

\newcommand{\HF}{{\rm HF}}

\newcommand{\SU}{{\rm SU}}

\newcommand{\nospin}{\text{\textnormal{no-spin}}}

\renewcommand{\epsilon}{\varepsilon}

\newcommand{\norm}[1]{ \left| \! \left| #1 \right| \! \right| }

\author{David Gontier}
\email{gontier@ceremade.dauphine.fr}
\address{CEREMADE, University of Paris-Dauphine, PSL University, 75016 Paris, France}

\author{Mathieu Lewin}
\email{Mathieu.Lewin@math.cnrs.fr}
\address{CNRS and CEREMADE, University of Paris-Dauphine, PSL University, 75016 Paris, France}

\title[Spin symmetry breaking in the HF electron gas]{Spin symmetry breaking in the translation-invariant Hartree-Fock electron gas}

\date{\today}

\begin{document}
\maketitle

\begin{abstract}
We study the breaking of spin symmetry for the nonlinear Hartree-Fock model describing an infinite translation-invariant interacting quantum gas (fluid phase). At zero temperature and for the Coulomb interaction in three space dimensions, we can prove the existence of a unique first order transition between a pure ferromagnetic phase at low density and a paramagnetic phase at high density. Multiple first or second order transitions can happen for other interaction potentials, as we illustrate on some examples. At positive temperature $T>0$ we compute numerically the phase diagram in the Coulomb case. We find the paramagnetic phase at high temperature or high density and a region where the system is ferromagnetic. We prove that the equilibrium state is unique and paramagnetic at high temperature or high density.
\end{abstract}

\bigskip

In this article, we study the Hartree-Fock (HF) model for an infinite interacting quantum Fermi gas, restricting our attention to fully translation-invariant states  (fluid phase). This model is solely parametrised by the density $\rho>0$ and the temperature $T\geq0$ of the gas, and it can be written as a minimisation problem over fermionic translation-invariant one-particle density matrices. Mathematically, we obtain a nonlinear minimisation problem involving a matrix-valued function $\gamma(\bk)$, where $\bk$ is the Fourier variable, which satisfies $0\leq \gamma(\bk)=\gamma(\bk)^*\leq 1$ in the sense of $2\times2$ hermitian matrices and the constraint that $\int_{\R^d}\tr_{\C^2}\gamma(\bk)\,{\rm d}\bk=(2\pi)^d\rho$. 

In three space dimensions and with the Coulomb interaction,
we obtain the Uniform Electron Gas (UEG). This is the reference model in Density Functional Theory~\cite{ParYan-94,PerKur-03}, where it appears in the Local Density Approximation~\cite{HohKoh-64,KohSha-65,LewLieSei-18} and is used for deriving the most efficient empirical functionals~\cite{Perdew-91,PerWan-92,Becke-93,PerBurErn-96,SunPerRuz-15,Perdew_etal-16}. Valence electrons in alcaline metals have indeed been found to be described by this model to a high precision, for instance in solid sodium~\cite{Huotari-etal-10}. Of course, the true ground state of the UEG is highly correlated at low and intermediate densities and Hartree-Fock theory provides a very rough approximation. But understanding the behaviour of mean-field theory is an important first step before developing more complicated methods including correlation.

There has been a huge recent interest in understanding the breaking of \emph{translational symmetry} in Hartree-Fock UEG~\cite{ZhaCep-08,BerDelHolBag-11,BagDelBerHol-13,BagDelBerHol-14,Baguet-14,GonHaiLew-19}. The corresponding phase diagram is very rich and a strong activity is currently devoted to exploring its properties in detail. Here we focus on the breaking of \emph{spin symmetry} and assume translation-invariance throughout, which is a much easier situation. In physical terms, we investigate the phase diagram of the fluid phase of the Hartree-Fock gas.

It is a well-established fact mentioned in many Physics textbooks~\cite{ParYan-94,GiuVig-05} that, at zero temperature, the translation-invariant system undergoes a first-order phase transition from a ferromagnetic phase with all spins aligned in one direction at low densities, to a paramagnetic phase with all spins independent from each other at high densities. However, the argument for this phenomenon is often reduced to comparing the energies of these two states, without actually showing that these are the only possible minimisers. In this paper, we provide the missing rigorous argument and give a complete proof of this phase transition. This phenomenon is however specific to the Coulomb potential. Multiple first or second order phase transitions can happen for other interaction potentials, as we illustrate on some examples.

In this work we also address the positive temperature case, which is much more involved and which we cannot solve completely. We compute numerically the full phase diagram for the Coulomb interaction in 3D and find two regions: the pure paramagnetic phase at high temperature or high density, and a region where the system is ferromagnetic. The transition from the paramagnetic phase to the ferromagnetic phase can be first or second order, depending on the values of $T$ and $\rho$. We are able to rigorously prove that the equilibrium state is unique and paramagnetic at high temperature or high density, but cannot rigorously justify the whole phase diagram. 

Mathematically, the problem is reduced to studying a nonlinear integral equation for matrix-valued radial functions. This nonlinear equation involves an Euler-Lagrange multiplier $\mu$ (called the chemical potential), associated with the constraint on the density $\rho$. We emphasise that solutions are in general \emph{non unique for fixed $\mu$}, a situation which is for instance different from the well-known case of the nonlinear Schr\"odinger equation~\cite{Tao-06}. This is the deep reason for the breaking of spin symmetry, as we explain in detail later. In the model we study, it is the competition between the (concave) exchange term and the (convex) entropy term which is responsible for this non uniqueness. Similar effects have been found recently for instance for the Bogoliubov model describing an infinite translation-invariant Bose gas but the phase transition is there due to the interplay between pairing and Bose-Einstein condensation~\cite{NapReuSol-18a,NapReuSol-18b,NapReuSol-18c}. 

\medskip

\subsubsection*{Acknowledgments.} The authors thank Majdouline Borji who contributed to the proof of Theorem~\ref{th:uniqueness_T=0} at $T=0$ during an internship at the University Paris-Dauphine in the summer 2017. They also thank Christian Hainzl with whom they found the estimate proved in~\cite{GonHaiLew-19} during the preparation of this work, which turned out to be useful for the proof of Theorem~\ref{th:uniqueness_positive_T}. This project has received funding from the European Research Council (ERC) under the European Union's Horizon 2020 research and innovation programme (grant agreement MDFT No 725528 of M.L.).

\section{Translation-invariant Hartree-Fock model}

\subsection{Hartree-Fock (free) energy}

We consider spin-polarised translation-invariant Hartree-Fock states in an arbitrary space dimension $d\geq1$. These are fully described by their one-particle density matrix which, in Fourier space, is a $\bk$-dependent $2\times2$ hermitian matrix
$$
\gamma(\bk):=
\begin{pmatrix}
\gamma^{\uparrow \uparrow}(\bk) & \gamma^{\uparrow \downarrow}(\bk)\\
\gamma^{\downarrow \uparrow}(\bk) & \gamma^{\downarrow \downarrow}(\bk)
\end{pmatrix},\qquad \bk\in\R^d.
$$
The Pauli principle is expressed by the condition that 
$$0\leq \gamma(\bk)\leq \bbI_2,$$
pointwise in the sense of $2\times2$ hermitian matrices. The corresponding density is the constant given by
$$\rho_\gamma:=\frac{1}{(2\pi)^d}\int_{\R^d}\tr_{\C^2}\gamma(\bk)\,\rd \bk,$$
where $\tr_{\C^2} \gamma(\bk) := \gamma^{\spinup \spinup}(\bk) + \gamma^{\spindown \spindown}(\bk)$
denotes the usual trace of $2 \times 2$ matrices.

We assume that the fermions interact through a repulsive radial potential $\check{w}$ to be chosen later on. The Hartree-Fock energy per unit volume of $\gamma$ is then given by
\begin{align}
\cE^{\rm HF}(\gamma) &:=
 \frac12 \frac{1}{(2 \pi)^d} \int_{\R^d} |k|^2\, \tr_{\mathbb{C}^2}\gamma(\bk)\, \rd \bk
- \frac12  \int_{\R^d} \check{w}(\bx)  \tr_{\C^2} | \hat{\gamma}(x) |^2 \rd \bx\nn\\
&=\frac{1}{2(2 \pi)^d} \int_{\R^d} |k|^2 \tr_{\mathbb{C}^2}\gamma(\bk) \rd \bk\nn \\
&\qquad\qquad- \dfrac{1}{2(2 \pi)^{d}} \iint_{\R^d\times\R^d} w(\bk - \bk') \tr_{\C^2} \left[ \gamma (\bk) \gamma(\bk') \right] \rd \bk\, \rd \bk'.
\label{eq:HF_energy}
\end{align}
The second term is the \emph{exchange energy} and it can be either expressed in terms of the translation-invariant kernel $\hat{\gamma}(x-y)$ in space of the Fourier multiplier $\gamma(\bk)$ (first equality) or in the Fourier domain (second equality). Here and in the sequel, $w(k)$ is the Fourier transform (up to a $(2 \pi)^{d/2}$ factor) of the interaction potential. The direct (or Hartree) term has been dropped, since it only depends on the constant $\rho_\gamma$ and plays no role in the minimisation. For potentials $\check{w}\notin L^1(\R^d)$ such as the Coulomb potential, the Hartree term is removed by the addition of a uniform background of positive charge.

At $T=0$ we are interested in minimising the HF energy per unit volume~\eqref{eq:HF_energy} over all possible states with given density $\rho_\gamma=\rho$
\begin{equation}
 \boxed{E^{\rm HF}(\rho) :=\min \left\{\cE^{\rm HF}(\gamma), \ 0\leq \gamma=\gamma^*\leq \bbI_2, \ \frac1{(2\pi)^{d}}\int_{\R^d} \tr_{\C^2} \gamma=\rho \right\}.}
 \label{eq:GS_HF_energy}
\end{equation}
We also would like to determine the form of the corresponding minimisers, depending of the value of $\rho$. 

At positive temperature $T>0$, we have to minimise the free energy per unit volume, which is given by
\begin{equation}
\cE^\HF(\gamma, T) :=  \cE^\HF(\gamma)-\frac{T}{(2 \pi)^d} \int_{\R^d} \tr_{\mathbb{C}^2} S(\gamma(\bk)) \, \rd \bk,
\label{eq:HF_free_energy}
\end{equation}
where $S(t) :=- t \log(t) - (1 - t) \log(1 - t)$ is the usual (concave) Fermi-Dirac entropy. The corresponding minimal free energy is
\begin{equation}
\boxed{E^{\rm HF}(\rho,T) :=\min \left\{\cE^{\rm HF}(\gamma, T), \ 0\leq \gamma=\gamma^*\leq \bbI_2, \ \frac1{(2\pi)^{d}}\int_{\R^d} \tr_{\C^2} \gamma=\rho \right\}.}
\label{eq:GS_HF_free_energy}
\end{equation}

\subsection{Spin symmetric states and the no-spin free energy}

A \emph{pure ferromagnetic HF state} has all its spins aligned in one direction and this corresponds to taking a density matrix in the form
$$\gamma_{\rm ferro}(\bk)=U\begin{pmatrix}g_{\rm ferro}(\bk)&0\\ 0&0\\ \end{pmatrix}U^*=g_{\rm ferro}(\bk)\, |Ue_1\rangle\langle Ue_1|$$
where the unitary $U\in \SU(2)$ determines the common polarisation $\nu=Ue_1$ of the spins. Its free energy is independent of $U$.
A \emph{paramagnetic HF state} has its spins chosen at random independently, with the uniform measure over all directions, and this corresponds to taking 
$$\gamma_{\rm para}(\bk)=g_{\rm para}(\bk)\,\bbI_2.$$
A \emph{general ferromagnetic HF state} is a non-trivial convex combination of a pure ferromagnetic state and a paramagnetic state, that is, a state of the form
$$\gamma(\bk)=U\begin{pmatrix}g_\spinup(\bk)&0\\ 0&g_{\spindown}(k)\\ \end{pmatrix}U^*$$
with $g_\spinup\neq g_\spindown$ and $g_{\spinup,\spindown}\neq0$. 

Because of the special form of these states, it is natural to introduce the no-spin version of the free energy~\eqref{eq:GS_HF_free_energy}, given by
\begin{multline}
\cE_\nospin^\HF(g, T) :=   \frac{1}{2(2 \pi)^d} \int_{\R^d} |k|^2 g(\bk) \rd \bk\\
- \dfrac{1}{2(2 \pi)^{d}} \iint_{\R^d\times\R^d} w(\bk - \bk') g(\bk) g(\bk')  \rd \bk\, \rd \bk'  -\frac{T}{(2 \pi)^d} \int_{\R^d} S(g(\bk)) \rd \bk,
\label{eq:HF_free_energy_no_spin}
\end{multline}
as well as the corresponding free energy
\begin{equation}
\boxed{ E_\nospin^{\rm HF}(\rho,T):=\min \left\{ \cE_\nospin^{\rm HF}(g,T), \ 0 \le g \le 1, \  \frac1{(2\pi)^{d}}\int_{\R^d}g(\bk)\,\rd\bk=\rho \right\}.}
\label{eq:GS_HF_free_energy_no_spin}
\end{equation}
Here $g$ is now a real-valued function. Similarly to the spin-polarised case, we use the simpler notation $\cE_\nospin^{\rm HF}(g):= \cE_\nospin^{\rm HF}(g,0)$ and $E_\nospin^{\rm HF}(\rho):=E_\nospin^{\rm HF}(\rho,0)$ at zero temperature.

\section{Main results on HF equilibrium states and on the phase diagram}

In this section we state our main mathematical results on HF equilibrium states and on the phase diagram. For convenience, some of the proofs will be given later in Section~\ref{sec:proof_thm_positive_T} and Appendix~\ref{sec:proof_lemma_well_posed}.

\subsection{Existence of minimisers}

First we state the following elementary result concerning the existence of minimisers.

\begin{lemma}[Well-posedness and existence of minimisers]\label{lem:well-posed}
We assume that $w\in L^1(\R^d)+L^\ii(\R^d)$ and that $\rho,T \geq0$. Then the (spin and no-spin) minimisation problems~\eqref{eq:GS_HF_free_energy} and~\eqref{eq:GS_HF_free_energy_no_spin} are well-posed and have minimisers. 

At $T>0$, any minimiser for~\eqref{eq:GS_HF_free_energy} solves the nonlinear equation
    \begin{equation} \label{eq:EulerLagrange_gamma}
    \gamma(\bk)=\left(1+\re^{\beta\big(\frac{k^2}{2} - \gamma * w(\bk)-\mu \big)}\right)^{-1},
    \end{equation}
for some $\mu \in \R$ called the chemical potential and with $\beta=1/T<\ii$. In particular, $0<\gamma(k)<\bbI_2$ for all $k\in\R^d$, in the sense of $2\times2$ hermitian matrices.

At $T=0$, any minimiser for~\eqref{eq:GS_HF_energy} solves the nonlinear equation
    \begin{equation} \label{eq:EulerLagrange_T0_gamma}
    \gamma(\bk)=\1\left(\frac{k^2}{2} - \gamma\ast w(\bk)< \mu\right)+\tilde\gamma(k)
    \end{equation}
    where $\tilde\gamma(k) \subset \ker\left(\frac{k^2}{2} - \gamma\ast w(\bk)- \mu\right)$ for every $k\in\R^d$. 

    Similar equations hold for the no-spin minimisers of~\eqref{eq:GS_HF_free_energy_no_spin}. In this case, if $w$ is radial non-increasing, then so are all the minimisers $g$. 
\end{lemma}

The result follows from classical methods in the calculus of variations, using that $|k|^2\to\infty$ at infinity. When $w$ is radial decreasing, then $g$ satisfies the same property by usual symmetric rearrangement inequalities for functions~\cite{LieLos-01}. The detailed proof of Lemma~\ref{lem:well-posed} is provided for completeness later in Appendix~\ref{sec:proof_lemma_well_posed}.

\subsection{Reduction to the no-spin problem}

Our main first result concerns the form of minimisers of~\eqref{eq:GS_HF_energy} and~\eqref{eq:GS_HF_free_energy} and the link with the no-spin counterpart~\eqref{eq:HF_free_energy_no_spin}. We assume here that $w$ is positive and recall that $w$ is the Fourier transform of the interaction potential.

\begin{theorem}[HF equilibrium states] \label{thm:spin_as_nospin}
We assume that $w\in L^1(\R^d)+L^\ii(\R^d)$ is a positive function and that $\rho,T \ge 0$.
Then, the minimisers of the spin problem~\eqref{eq:GS_HF_free_energy} are all of the form
$$\gamma(\bk)=U\begin{pmatrix}g_\spinup(\bk)&0\\
0&g_\spindown(\bk)
\end{pmatrix}U^*$$
where $U\in \SU(2)$ is a $\bk$-independent unitary matrix and where $g_{\spinup}$ and $g_{\spindown}$ are minimisers of the no-spin problem~\eqref{eq:GS_HF_free_energy_no_spin}, for some densities $\rho^\spinup$ and $\rho^\spindown$ respectively (to be determined and satisfying $\rho^\spinup + \rho^\spindown = \rho$). In particular, 
\begin{align}
E^{\rm HF}(\rho,T)&=\min_{\rho^\spinup+\rho^{\spindown}=\rho}\big\{E_\nospin^{\rm HF}(\rho^\spinup,T)+E_\nospin^{\rm HF}(\rho^\spindown,T)\big\}\nn\\
&=\min_{0\leq t\leq 1/2}\Big\{E_\nospin ^{\rm HF}\big(t\rho,T\big)+E_\nospin^{\rm HF}\big((1-t)\rho,T\big)\Big\}. \label{eq:decoupled_energy}
\end{align}
\end{theorem}

This result states that minimisers at density $\rho\geq0$ and temperature $T\geq0$ are always made of $t\rho$ spins pointing in one fixed ($k$-independent) direction and $(1-t)\rho$ spins pointing in the other direction, with both density matrices minimising the corresponding no-spin problems. Here $0\leq t\leq1/2$ is a mixing parameter to be determined, called the \emph{polarisation}, with $t=1/2$ corresponding to the paramagnetic phase and $t=0$ to the pure ferromagnetic phase. For $0\leq t<1/2$ the material has a non trivial (partial) polarisation, and is also called ferromagnetic. 

In the model with spin, minimisers can be unique only in the paramagnetic case $t=1/2$. This is because otherwise we can rotate the spins as we like by applying a $U\in {\rm SU}(2)$. The non-uniqueness of minimisers is the manifestation of spin symmetry breaking. 

Note that the \emph{pure} ferromagnetic state $t=0$ can never occur at $T>0$. Indeed, as stated in Lemma~\ref{lem:well-posed}, $\gamma(k)$ can never have 0 as eigenvalue. In particular, the optimal $t$ in~\eqref{eq:decoupled_energy} is always positive at $T>0$.

\begin{remark}[Linear response to an external magnetic field]
When a constant magnetic field $B$ is applied to the system, we expect the energy per unit volume to behave as $\frac{1}{2(2 \pi)^d} (\frac12 -t) | B |$ to first-order. If the system is paramagnetic ($t = \frac12$), the energy behaves quadratically in $|B|$, while if the system is ferromagnetic ($t < \frac12$), the spins align along the direction of $B$, and the energy decreases linearly with $| B |$.
\end{remark}

\medskip

The proof of Theorem~\ref{thm:spin_as_nospin} relies on a kind of rearrangement inequality for matrices (Lemma~\ref{lem:rearrangement_matrices} below), which states that $\tr(UD_1U^*D_2)\leq \tr(D_1D_2)$ for all $U\in{\rm SU}(2)$ and any two diagonal positive matrices $D_1$ and $D_2$ with entries ordered in the same manner. Together with the positivity of $w$, this allows to show that the exchange term favours having $\gamma(k)$ diagonalised in a $k$-independent basis. 

\begin{proof}[Proof of Theorem~\ref{thm:spin_as_nospin}] 
Let $\gamma$ be any minimiser for~\eqref{eq:GS_HF_free_energy}. We may write
$$\gamma(\bk)=U(\bk)\begin{pmatrix}g_\spinup(\bk)&0\\
0&g_\spindown(\bk)
\end{pmatrix}U(\bk)^*$$
with for instance $g_\spinup(\bk)\geq g_\spindown(\bk)$. We then claim that $\cE^{\rm HF}(\tilde \gamma, T)\leq \cE^{\rm HF}(\gamma, T)$ for the diagonal state
$$\tilde \gamma(\bk)=\begin{pmatrix}g_\spinup(\bk)&0\\
0&g_\spindown(\bk)
\end{pmatrix},$$
that is, the energy goes down by decoupling the two spin states. Since the kinetic energy and the entropy are unchanged, we only have to explain why the exchange energy decreases. This follows from the next lemma, which is valid in any dimension but which we state for simplicity for $2\times2$ matrices.

\begin{lemma}[Rearrangement inequality for matrices] \label{lem:rearrangement_matrices}
    Let 
    $$D_1 = \begin{pmatrix}\lambda_1 & 0\\0 & \mu_1 \end{pmatrix},\qquad D_2=\begin{pmatrix}\lambda_2 & 0\\0 & \mu_2
    \end{pmatrix}$$ 
    be two diagonal matrices with eigenvalues ordered as $\lambda_1\geq\mu_1$ and $\lambda_2\geq\mu_2$. Then, for any unitary matrix $U \in \SU(2)$, we have 
    $$\tr_{\C^2}(D_1 U D_2 U^{*}) \leq \tr_{\C^2}(D_1D_2),$$
    with equality if and only if $U D_1 U^* = D_1$ or $U D_2 U^* = D_2$. If furthermore $\lambda_1>\mu_1$ and $\lambda_2>\mu_2$, then there is equality if and only if $U$ is diagonal. 
\end{lemma}

\begin{proof}
    We denote by $J := \begin{pmatrix}1 & 0\\0 & 0\end{pmatrix}$ and by $\alpha_i := (\lambda_i - \mu_i) \ge 0$, so that $D_i=\mu_{i}+\alpha_{i}J$. We get
    \[
    \tr_{\C^2}(D_1D_2)  - \tr_{\C^2}(D_1 U D_2 U^{*})= \alpha_{1} \alpha_{2}\left[ 1 - \tr_{\C^2}(JUJU^{*}) \right] =  \alpha_{1} \alpha_{2}\left[ 1 - | U_{11} |^2 \right] \ge 0,
    \]
    where we have used the fact that $| U_{11} | \le 1$ for any $U \in \SU(2)$. We have equality if and only if $\alpha_i = 0$ (in which case $D_i$ is a multiple of the identity and $U D_i U^* = D_i$), or if $| U_{11} | = 1$ (in which case $U$ is diagonal, and then $U D_{1,2} U^* = D_{1,2}$).
\end{proof}

Applying the lemma to the exchange energy, using that $w\geq0$, gives, as we wanted, that
 \begin{align*}
& \iint_{\R^d\times\R^d} w(\bk - \bk') \tr_{\C^2} \left[ \gamma (\bk) \gamma(\bk') \right] \rd \bk\, \rd \bk'\\
&\qquad \leq \iint_{\R^d\times\R^d} w(\bk - \bk') \tr_{\C^2} \left[ \tilde \gamma (\bk) \tilde \gamma(\bk') \right] \rd \bk\, \rd \bk'\\
 &\qquad =\iint_{\R^d\times\R^d} w(\bk - \bk')  g_\spinup(\bk)g_\spinup(\bk') \rd \bk\, \rd \bk'+ \iint_{\R^d\times\R^d} w(\bk - \bk')  g_\spindown(\bk)g_\spindown(\bk') \rd \bk\, \rd \bk'.
 \end{align*}
In particular, we find
$$\cE^{\rm HF}(\gamma, T)\geq \cE_\nospin^{\rm HF}(g_\spinup, T)+\cE_\nospin^{\rm HF}(g_\spindown, T) \geq  E^{\rm HF}_\nospin(\rho_\spinup,T)+E_\nospin^{\rm HF}(\rho_\spindown,T).$$
Since the reverse inequality can be obtained by taking diagonal trial states, we conclude that~\eqref{eq:decoupled_energy} holds. In addition, $g_\spinup$ and $g_\spindown$ minimise $E_\nospin^{\rm HF}(\rho_\spinup,T)$ and $E_\nospin^{\rm HF}(\rho_\spindown,T)$, respectively.

It remains to explain that the unitary $U(\bk)$ can indeed be chosen independent of $\bk$. If $\gamma(k)$ is a multiple of the identity, it is obvious that $U(k)\gamma(k)U(k)^*=\gamma(k)$ for any $U(k)\in {\rm SU}(2)$, so we can remove the unitary. We have to prove the similar equality in the region where $\gamma(k)$ is not a multiple of the identity. Let $k'$ be in this region. Then, for every other point $k$ in the same region we have from  the assumed positivity of $w$ that
$$\tr_{\C^2}\left[U(\bk)\tilde\gamma(\bk)U(\bk)^*U(\bk')\tilde\gamma(\bk')U(\bk')^*\right]=\tr_{\C^2}\left[\tilde\gamma(\bk)\tilde\gamma(\bk')\right].$$
According to Lemma~\ref{lem:well-posed} this implies that $U(\bk)^*U(\bk')$ is a diagonal matrix for all such $\bk,\bk'$. Therefore, $U(\bk)$ is equal to the fixed unitary $U=U(k')$ times a diagonal unitary matrix, commuting with $\gamma(k)$. This proves that $\gamma(\bk)=U\tilde\gamma(\bk) U^*$ and concludes the proof of Theorem~\ref{thm:spin_as_nospin}.
\end{proof}

\subsection{Uniqueness and non-uniqueness for the no-spin model}

In order to find the optimal value of $t\in[0,1/2]$  for the minimisation problem~\eqref{eq:decoupled_energy}, we need to discuss the no-spin minimisation problem with more details.

We know from Theorem~\ref{thm:spin_as_nospin} that, up a global unitary, any minimiser $\gamma$ takes the special form
$$\gamma(k)=\begin{pmatrix}g_\uparrow(k)&0\\
0&g_\downarrow(k)
\end{pmatrix}$$
with $g_{\uparrow,\downarrow}$ minimising the no-spin free energy for the corresponding (unknown) $\rho_{\uparrow,\downarrow}$, hence solving the corresponding nonlinear equation. Since $\gamma$ should satisfy an equation similar to that of $g_{\uparrow,\downarrow}$, the Lagrange multipliers of $g_{\uparrow,\downarrow}$ must be the same:
$$\mu_\uparrow=\mu_\downarrow=\mu.$$
From this property we see that we can have spin symmetry breaking, that is $t<1/2$, only when there are two no-spin minimisers $g_\uparrow$ and $g_\downarrow$ with different total densities, $\rho_\uparrow\neq \rho_\downarrow$, but sharing the same chemical potential $\mu_\uparrow=\mu_\downarrow$. In particular, spin-symmetry breaking can only happen if there is \emph{non-uniqueness} of solutions to the equation
\begin{equation} \label{eq:EulerLagrange_mu}
    g(\bk)=\left(1+\re^{\beta\big(\frac{k^2}{2} - g \ast w(\bk)-\mu\big)}\right)^{-1},
    \end{equation}
at fixed chemical potential $\mu\in\R$, or the equivalent equation 
    \begin{equation} \label{eq:EulerLagrange_mu_T0}
    g(\bk)=\1\left(\frac{k^2}{2} - g\ast w(\bk)\leq \mu \right)+\tilde g(k),
    \end{equation}
at $T=0$. Note that such solutions exist for all $\mu\in\R$, as is seen by minimising the free energy $\cE^{\rm HF}_\nospin(T,g)-\mu\rho_g$ without the density constraint. 

Even though equilibrium states of the no-spin problem are in general not unique at fixed chemical potential $\mu$, we believe that they are unique when parametrised in terms of the density $\rho\geq0$.

\begin{conjecture}[Uniqueness for the no-spin problem] \label{conj:uniqueness}
When $w$ is a positive radial non-increasing function, the no-spin minimisation problem $E_\nospin^{\rm HF}(\rho,T)$ in~\eqref{eq:GS_HF_free_energy_no_spin} admits a unique minimiser for every $\rho,T\geq0$.
\end{conjecture}

One traditional argument, often used in the study of Partial Differential Equations, is to prove the uniqueness of (radial) solutions of the equations~\eqref{eq:EulerLagrange_mu} and~\eqref{eq:EulerLagrange_mu_T0} for any given $\mu$, which then implies the uniqueness of minimisers. As explained, this will not work here and this complicates the mathematical analysis.

\subsection{Uniqueness at zero temperature}

At zero temperature we are able to solve Conjecture~\ref{conj:uniqueness} completely. We prove the uniqueness of minimisers of the no-spin problem for all possible values of the density $\rho$, under the additional assumption that $w$ is radial non-increasing, which in particular covers the Coulomb case. 

\begin{theorem}[Uniqueness at $T = 0$] \label{th:uniqueness_T=0}
We assume that $w\in L^1(\R^d)+L^\ii(\R^d)$ is a positive radial non-increasing function. Then, for $T=0$ and all $\rho\geq 0$, the no-spin minimisation problem~\eqref{eq:GS_HF_free_energy_no_spin} has a \emph{unique minimiser} $g_{\rho,0}$, given by
\begin{equation}
 g_{\rho,0}(\bk)=\1\Big(k^2\leq c_{\rm TF}\,\rho^{2/d}\Big),
 \quad \text{where} \quad
 c_{\rm TF}:=4\pi^2 \left( \dfrac{d}{| \SS^{d-1} |}\right)^{2/d}
 \label{eq:form_GS_T_0}
\end{equation}
is the Thomas-Fermi constant.
\end{theorem}

It may seem surprising at first sight that, at $T=0$, the no-spin ground state $g_{\rho, 0}$ is independent of $w$ and is given by the same formula as when $w\equiv0$. This is a consequence of the property that $w(\bk)$ is radial non-increasing. The argument is as follows.

\begin{proof}[Proof of Theorem~\ref{th:uniqueness_T=0}]
When the interaction potential $w$ is radial non-increasing, minimisers for the no-spin problem~\eqref{eq:GS_HF_free_energy_no_spin} are also radial non-increasing, by Lemma~\ref{lem:well-posed}. This implies that $g\ast w$ is radial non-increasing as well, hence that
$$k\mapsto  h(k)=\frac{k^2}{2}-g\ast w(\bk)$$
is radial and strictly increasing. Since $g$ satisfies the equation~\eqref{eq:EulerLagrange_mu_T0} for some $\mu$, it must then be the characteristic function of a ball. The radius of the ball is found from the constraint on the density.
Note that $\tilde g(k)$ vanishes since the level sets of the above function $h(k)$ are spheres, hence have vanishing Lebesgue measure.
\end{proof}

With Theorem~\ref{th:uniqueness_T=0} at hand we can compute exactly the right side of~\eqref{eq:decoupled_energy} and determine the optimal values of $t$ and the possible regions of symmetry breaking. This is done in Section~\ref{sec:T_0}, where the zero-temperature case is investigated for general Riesz-type potentials in all dimensions. In the three dimensional Coulomb case, we need to find the minimum of the function
\[
    P_\rho(t) := \left( \dfrac{3^{5/3} \pi^{4/3}}{2^{1/3} 5} \right) \rho^{5/3} (t^{5/3} + (1 - t)^{5/3}) - \left( \dfrac{3^{4/3}}{2^{5/3} \pi^{1/3}} \right) \rho^{4/3} (t^{4/3} + (1 - t)^{4/3}).
\]
In Section~\ref{sec:T_0} we prove the following result.

\begin{corollary}[First order phase transition for the 3D Coulomb case at $T=0$]\label{cor:transition_Coulomb_T0}
We assume that $w(k)=(2\pi^2)^{-1}|k|^{-2}$, $d=3$ and that $T=0$. Then we have a first-order phase transition between the ferromagnetic and paramagnetic phases at density
\begin{equation} \label{eq:numericalValueRhoc_pre}
\rho_c = \dfrac{125}{24 \pi^5} \left( \dfrac{1}{1+2^{1/3}}\right)^3 \approx 1.47 \times 10^{-3}.
\end{equation}
More precisely, 
    \begin{itemize}
        \item for all $0 < \rho < \rho_c$, the minimisers of $E^{\rm HF}(\rho)$ are all of the form 
$$\gamma_{\rm ferro}(\bk)=\1\big(k^2\leq c_{\rm TF}\rho^{2/3}\big)\;|\nu\rangle\langle\nu| \qquad \text{(ferromagnetic phase)}$$ 
with $\nu$ any normalised vector in $\C^2$;
        \item for all $\rho > \rho_c$, the minimiser of $E^{\rm HF}(\rho)$ is unique, and given by
        $$\gamma_{\rm para}(\bk) := \1\big(k^2\leq c_{\rm TF} (\rho/2)^{2/3}\big)\, \bbI_2 \qquad \text{(paramagnetic phase)};$$
        \item for $\rho = \rho_c$, the minimisers are of either form.
    \end{itemize}
The first derivative of the ground state energy $\rho\mapsto E^{\rm HF}(\rho)$ has a jump at $\rho=\rho_c$. 
    \end{corollary}

\begin{remark}
    The critical $\rho_c$ found in~\eqref{eq:numericalValueRhoc_pre} gives, in terms of the Wigner-Seitz radius,
    $$r_s := \left( \frac{3}{4 \pi \rho_c} \right)^{1/3} \approx 5.45.$$ 
    This is the result found in standard Physics textbooks (see e.g.~\cite[Eqt. 2.57]{GiuVig-05}).
\end{remark}

\subsection{Phase diagram in the 3D Coulomb case}

\begin{figure}[H]
    \centering
         \includegraphics[width=13cm]{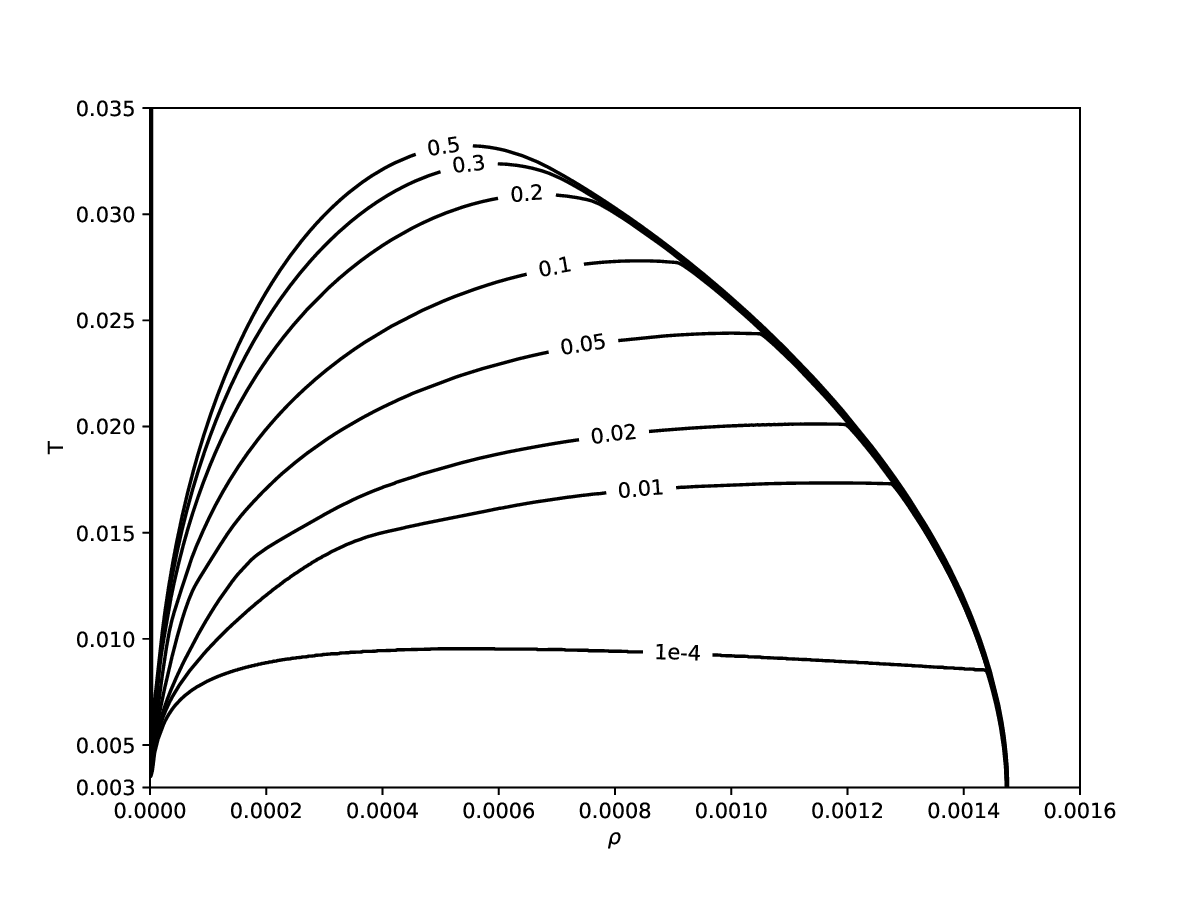}
    \caption{Phase diagram for the Hartree-Fock gas. The $0.5$ region corresponds to paramagnetism.}
    \label{fig:phaseDiagram}
\end{figure}

We next turn to the positive temperature case, which we cannot solve completely. 
In Figure~\ref{fig:phaseDiagram}, we display a numerical simulation of the polarisation $t$ of the electron gas, in the Coulomb case $d = 3$ and $w(k) = (2 \pi^2)^{-1} | k |^{-2}$, as a function of $\rho$ and $T$. We represent there the level sets of the polarisation ($0.5$ corresponds to paramagnetism, while $0$ corresponds to pure ferromagnetism). We took temperatures\footnote{The $T < 0.003$ region is sensitive to numerical noise, so we decided not to represent it in this figure.} $T \in [0.003, 0.035]$, and densities $\rho \in [0, 0.0016]$.

In this figure we observe that there is a critical Curie temperature $T_c \approx 0.034$ above which the gas is always paramagnetic. For $0 < T < T_c$, we find two transitions when $\rho$ increases. The system is paramagnetic at low density, then undergoes a first-order transition to ferromagnetism as the density passes a first critical temperature $\rho_{c,1}(T)$, and finally becomes again suddenly paramagnetic at a second critical density $\rho_{c,2}(T)$. 

As explained above we only find the pure ferromagnetic state at $T=0$ and $\rho\leq \rho_c\approx 0.00147$, which  matches the result found in~\eqref{eq:numericalValueRhoc_pre}. However, the polarisation seems to decrease quite rapidly with $T$ if we fix $\rho<\rho_c$. It would be interesting to determine the dependence of the optimal $t$ in~\eqref{eq:decoupled_energy} as we increase the temperature starting from the ferromagnetic state. 

Our method for computing the phase diagram relies on some tools introduced in the proof of Theorem~\ref{th:uniqueness_positive_T} which we are going to state in the next section. For this reason our numerical technique is quickly explained later in Remark~\ref{rmk:numerics}.

We have largely insisted on the fact that the breaking of spin symmetry is deeply related to the non-uniqueness of solutions to the equation~\eqref{eq:EulerLagrange_mu}, for given $\mu$. We quickly illustrate this now.

\begin{figure}[H]
    \includegraphics[width=10cm]{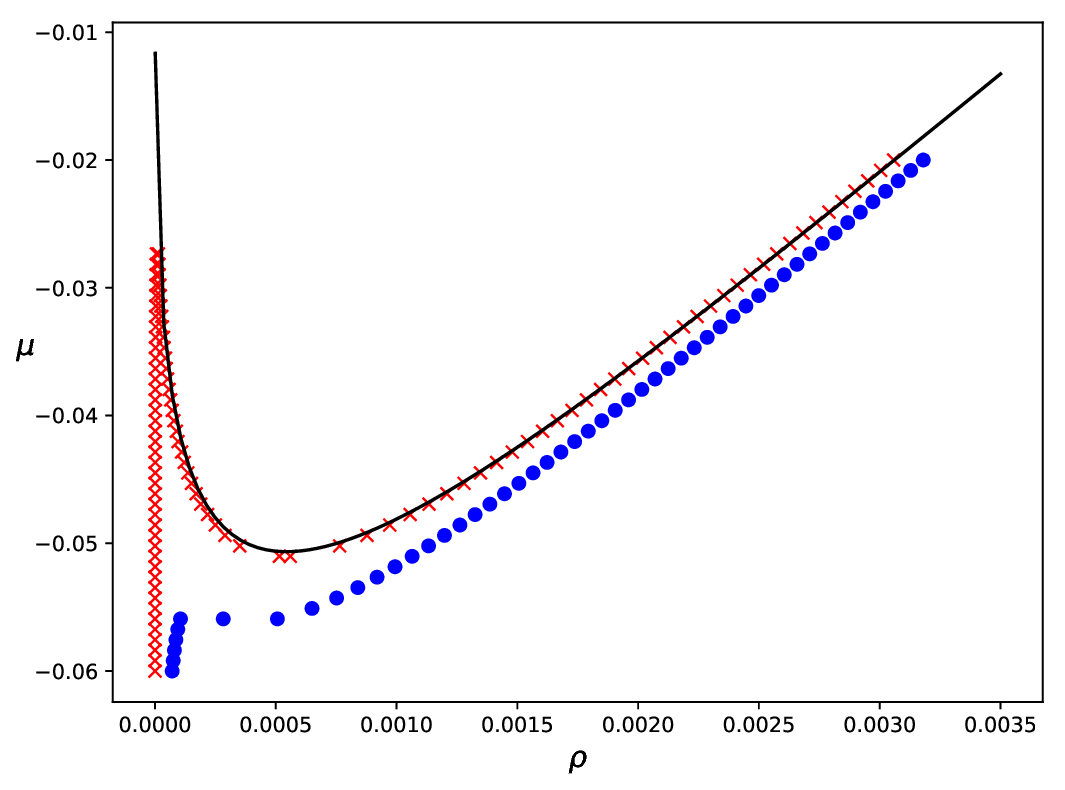}
    \caption{The Fermi level $\mu$ as a function of the density $\rho$ for the three-dimensional Coulomb gas, for different temperatures. Solid black line is $T = 0$ (see~\eqref{eq:mu_rho_T=0}), the x/red points represent $T = 0.01$, and the o/blue points represent $T = 0.03$.}
    \label{fig:rho_mu}
\end{figure}

At zero temperature, for the three-dimensional Coulomb case, we have from Equation~\eqref{eq:EulerLagrange_T0_gamma} and Theorem~\ref{th:uniqueness_T=0} that the Lagrange multiplier for the no-spin problem is given by (this is also the derivative of the energy with respect to the density, as we will prove later)
\begin{equation} \label{eq:mu_rho_T=0}
\mu(\rho, T = 0) =  \frac{3^{2/3}\pi^{4/3}}{2^{1/3}} \rho^{2/3} - \frac{6^{1/3}}{\pi^{1/3}} \rho^{1/3},
\end{equation}
which is not one-to-one. We therefore expect that $\rho \mapsto \mu(\rho, T)$ is also not one-to-one for small positive temperatures. This is confirmed by the numerical calculation displayed in Figure~\ref{fig:rho_mu}. There we plot in the $(\mu, \rho)$-plane the set of points for which the nonlinear equation~\eqref{eq:EulerLagrange_mu} with chemical potential $\mu$ admits a solution with density $\rho$, for different temperatures. We see that different values of the density $\rho$ may lead to the same Fermi level $\mu$, depending on the value of $T$. 

In the present figure, we notice several important features, that we use later in the proof of Theorem~\ref{th:uniqueness_positive_T}. For $T > 0$, there is $0 < \rho_1(T) \le \rho_2(T)$ such that the map $\rho \mapsto \mu(\rho, T)$ is strictly increasing on $[0,\rho_1(T)]$ and on $[\rho_2(T),\ii)$. Although we see that $\rho_1(T) \to 0$ as $T \to 0$, we are able to control this convergence, and we prove that $\rho_1(T) \ge C T^3$. On the other hand, when the temperature is high-enough, we have $\rho_1(T)=\rho_2(T)$ and the map $\rho \mapsto \mu(\rho, T)$ becomes increasing (hence one-to-one) over the whole of $\R^+$.

Although we believe that most of the numerical findings of this section apply to more general situations, we have not investigated other potentials nor other dimensions.

\subsection{Uniqueness at positive temperature}

In this section, we discuss uniqueness for the no-spin minimisation problem $E^{\HF}_{\nospin} (\rho, T)$ (see Conjecture~\ref{conj:uniqueness}) and for the spin-polarised one $E^{\HF} (\rho, T)$, in the case where $w(k)$ can be bounded by some $|k|^{s-d}$ in an appropriate manner. We are able to prove that both problems have a unique minimiser at large $T$ or large $\rho$. In particular, the spin problem $E^{\rm HF}(\rho, T)$ is paramagnetic in an appropriate region. 

 We first state our theorem for the no-spin problem. In our study, the potential $|k|^{1-d}$ (corresponding to $\check{w}(x) = C | x |^{-1}$) is critical in any dimension, as it is for the scattering of Schr\"odinger operators~\cite{ReeSim3}. We therefore split our theorem into two parts. The first deals with sub-critical potentials, whereas the second is the critical case $|k|^{1-d}$.

\begin{theorem}[Uniqueness for the no-spin problem at $T > 0$] \label{th:uniqueness_positive_T} ~\\
\noindent $\bullet$ \textsl{(Short-range potentials).} Let $w\in L^1(\R^d)+L^\ii(\R^d)$ be such that $w$ is positive radial non-increasing and satisfies the pointwise bound
\begin{equation}
 0< w(k) \leq \dfrac{\kappa_1}{|k|^{d-s_1}} +  \dfrac{\kappa_2}{|k|^{d-s_2}},
 \label{eq:bound_hat_w}
\end{equation}
with $1 < s_1 \le s_2 < 2$, and $\kappa_1,\kappa_2 \geq 0$. Then there exists $C, \rho_C \geq0$ such that, for all $(\rho,T)$ in the region
\begin{equation} \label{eq:def:Omega}
\Omega := \left\{ T>C \rho^{{s_1}/{d}} \quad\text{or}\quad \rho>\rho_C \right\},
\end{equation}
the function $g\mapsto \cE^{\rm HF}_{\nospin}(g,T)$ has a \emph{unique critical point} $g_{\rho,T}$ of density $\rho$, which is therefore the \emph{unique minimiser} for $E^{\rm HF}_{\nospin}(\rho,T)$. The function $g_{\rho,T}$ is positive radial-decreasing. It is non-degenerate, in the sense that the linearised operator
\begin{equation}
\cL_{\rho, T}f:=\frac{T}{g_{\rho, T}(1-g_{\rho, T})}f - w \ast f
\label{eq:linearised operator}
\end{equation}
is positive-definite on $L^2(\R^d)$. The map $(\rho,T)\mapsto g_{\rho,T}\in L^1(\R^d)\cap L^\ii(\R^d)$ is real-analytic in $\Omega$ and the corresponding (unique) Lagrange multiplier $\mu(\rho,T)$ satisfies
\begin{equation}
 \mu(\rho,T)=\frac{\partial}{\partial\rho}E^{\rm HF}_\nospin(\rho, T).
 \label{eq:formula_mu}
\end{equation}
Finally, for any interval $T \times [\rho_1,\rho_2]$ in the region $\Omega$, the function $\rho\in[\rho_1,\rho_2]\mapsto \mu(\rho, T)$ is strictly increasing hence the energy $\rho\in[\rho_1,\rho_2]\mapsto E^{\rm HF}_\nospin(\rho, T)$ is strictly convex.

\smallskip

\noindent $\bullet$ \textsl{(Long range potential $s=1$).} In the case where 
\begin{equation}
w(k) = \frac{\kappa}{| k |^{d-1}} 
 \label{eq:bound_hat_w_critical}
\end{equation}
with $d > 1$, all the previous conclusions hold after replacing $\Omega$ in~\eqref{eq:def:Omega} by
\begin{equation} \label{eq:def:Omega_Coulomb}
\Omega := \left\{  T \ge C \rho^{1/{d}} \re^{-\alpha \rho^{1/{d}}} \right\},
\end{equation}
for some $C \in \R^+$ and $\alpha >0$.
\end{theorem}

For the spin-polarised problem, we have the following result.
\begin{theorem}[Uniqueness and paramagnetism for the spin problem]
    \label{th:uniqueness_paramagnetism_positive_T}
    With the same assumptions on $w$ as in Theorem~\ref{th:uniqueness_positive_T}, there is $C \in \R^+$ and $\alpha > 0$ such that, for all $(\rho, T) \in \widetilde{\Omega}$, where
    \[
        \widetilde{\Omega} := 
        \begin{cases}       
        \left\{ T > C \rho^{s_1/d} \re^{ - \alpha \rho^{1/d}}    \right\} & \text{in the short range case},\\[0.4cm]
         \left\{ T > C \rho^{1/d} \re^{ - \alpha \rho^{1/(2d)}}    \right\} & \text{in the long range case $s=1$},
        \end{cases}
    \]
    the spin-polarised minimisation problem $E^\HF(\rho, T)$ has a unique minimiser, which is paramagnetic, and given by
    \[
    \gamma_{\rm para}(\bk) = g_{\rho/2, T}(k) \bbI_2,
    \]
with $g_{\rho/2,T}$ the unique minimiser of the no-spin problem $E_\nospin^\HF(\rho/2,T)$.
\end{theorem}

\begin{remark}[Curie temperature]
    There is a temperature $T_C > 0$ for which $\R^+ \times [T_C, \infty) \subset \Omega$. So the corresponding system is always paramagnetic in this region. The minimal $T_C$ having this property is sometimes called the~{\rm Curie temperature}.
\end{remark}

Theorems~\ref{th:uniqueness_positive_T} and~\ref{th:uniqueness_paramagnetism_positive_T} are the most involved results of this paper. Their lengthy proofs are detailed later in Section~\ref{sec:proof_thm_positive_T}. For Theorem~\ref{th:uniqueness_positive_T}, the uniqueness of critical points relies on the fact that for any $T>0$ and $\mu\in\R$, the nonlinear equation
$$g=\frac{1}{\re^{\beta(k^2/2-w\ast g-\mu)}+1}$$
takes the form of a \emph{Hammerstein equation}~\cite{BroGup-69}, with an ordering-preserving  nonlinear operator. It particular, there is always a minimal solution $g_{\rm min}$ and a maximal solution $g_{\rm max}$, in the sense that any solution $g$ satisfies $g_{\rm min}\le g\le g_{\rm max}$ pointwise. In addition, by construction, the functions $g_{\rm min}$ and $g_{\rm max}$ are radial-decreasing. The main ingredient of the proof is that any radial-decreasing solution which has $(\rho,T)\in\Omega$ is necessarily non-degenerate, hence gives rise to a smooth branch of solutions in its neighbourhood, by the implicit function theorem. Applying this to $g_{\rm min/max}$, which are always radial-decreasing, we are able to extend the two branches to the whole domain $\Omega$. We then show that $g_{\rm min}=g_{\rm max}$ in the whole region, by studying the region of small densities, where we can prove the uniqueness of critical points.  At high densities, our arguments are inspired by our recent work~\cite{GonHaiLew-19} which is itself based on spectral techniques recently developed in the context of Bardeen-Cooper-Schrieffer theory in~\cite{FraHaiNabSei-07,HaiHamSeiSol-08,HaiSei-08b,FreHaiSei-12,HaiSei-16,HaiLos-17}. For Theorem~\ref{th:uniqueness_paramagnetism_positive_T}, we directly prove the uniqueness of minimisers, using some estimates derived in the proof of Theorem~\ref{th:uniqueness_positive_T}. 

\begin{remark}
   The set $\widetilde{\Omega}$ is a strict subset of the set $\Omega$ appearing in Theorem~\ref{th:uniqueness_positive_T}. We conjecture that all the results of Theorems~\ref{th:uniqueness_positive_T}--\ref{th:uniqueness_paramagnetism_positive_T} are valid in a region of the form $\Omega := \{ T>C \rho^{{s_1}/{d}} \ \text{or}\ \rho>\rho_C \}$, even in the long range case. 
\end{remark}

\section{Detailed study of the phase transition(s) at $T=0$}\label{sec:T_0}


In this section we study the minimisation problem over $t\in[0,1/2]$ in~\eqref{eq:decoupled_energy} using Theorem~\ref{th:uniqueness_T=0} that provides the form of the unique minimiser for the no-spin problem at $T=0$. 

\subsection{Riesz (power-law) interactions}
\label{ssec:Riesz}
We look at the special case of the Riesz interactions
\[
    \boxed{w(\bk) = \dfrac{c_{d,s}}{| \bk |^{d-s}}}
    \quad \text{with} \quad
    c_{d,s} := \dfrac{1}{(2 \pi)^{d/2}} \dfrac{2^{(d-s)/2}}{2^{s/2}} \dfrac{\Gamma \left(\frac{d-s}{2}\right)}{\Gamma \left(\frac{s}{2}\right)},
\]
where $0 < s < d$. The constant $c_{d,s}$ is chosen so that $w(\bk)$ corresponds to the Fourier transform (up to some $(2 \pi)^{d/2}$ factor) of the interaction $ | \bx |^{-s}$. In addition to the Coulomb case $s=1$ in dimension $d=3$ (where $c_{3,1} = 1/(2 \pi^2)$), several physical systems may be appropriately described by such purely repulsive power-law potentials, including for instance colloidal particles in charge-stabilised suspensions~\cite{PauAckWol-96,SenRic-99} or certain metals under extreme thermodynamic conditions~\cite{HooYouGro-72}. 

By plugging the zero-temperature solution $g(\bk)=\1(k^2\leq c_{\rm TF}\rho^{2/d})$ found in Theorem~\ref{th:uniqueness_T=0}, we find immediately after scaling that the no-spin ground state energy is equal to
\begin{equation} \label{eq:explicitI}
        \boxed{E_\nospin^{\rm HF}(\rho,0) =  \kappa(d) \rho^{1+\frac{d}{2}}
        - \lambda(d,s) \rho^{1+\frac{s}{d}}}
\end{equation}
where
\begin{equation*}
        \kappa(d) := \dfrac{ 2\pi^2 d}{(d+2)} \left( \dfrac{d}{| \SS^{d-1} |} \right)^{2/d} , \quad
        \lambda(d,s) :=\dfrac{1}{2 \pi^{\frac{d}{2} - s}} \left( \dfrac{d}{| \SS^{d-1} |} \right)^{\frac{d+s}{d}}
        \dfrac{\Gamma \left(\frac{d-s}{2}\right)}{\Gamma \left( \frac{s}{2}\right)} c_{\rm D}(d,s)
\end{equation*}
and with the Dirac-type constant
\[
    c_{\rm D}(d, s) := \iint_{\R^d\times\R^d} \dfrac{\1(| \bk | < 1) \1 (| \bk' | < 1) }{| \bk - \bk' |^s}\, \rd \bk\ \rd \bk'.
\]
There happens to be a change of behaviour depending whether $s$ is below or above $\min(2,d)$. 

\begin{theorem}[Sharp or smooth phase transition for Riesz interactions]\label{thm:T_0}
Let $d\geq1$ and assume that $w(\bk) = c_{d,s}| \bk |^{s-d}$ with $0<s<d$.

\medskip

\noindent \textbf{(First-order phase transition).} If $0<s<\min(2,d)$, then there is a first-order transition between a ferromagnetic and a paramagnetic phase at density
 \begin{equation}
          \rho_c := \left( \dfrac{\lambda(d,s)}{\kappa(d)} \dfrac{1 - 2^{-s/d}}{1 - 2^{-2/d}} \right)^{\frac{d}{2-s}}.
 \end{equation}
More precisely, 
    \begin{itemize}
        \item for all $0 < \rho < \rho_c$, the minimisers of $E^{\rm HF}(\rho)$ are all of the form 
$$\gamma_{\rm ferro}(\bk)=\1\big(k^2\leq c_{\rm TF}\rho^{2/d}\big)\;|\nu\rangle\langle\nu| \qquad \text{(ferromagnetic phase)}$$ 
with $\nu$ any normalised vector in $\C^2$;
        \item for all $\rho > \rho_c$, the minimiser of $E^{\rm HF}(\rho)$ is unique, and given by
        $$\gamma_{\rm para}(\bk) := \1\big(k^2\leq c_{\rm TF} (\rho/2)^{2/d}\big)\, \bbI_2 \qquad \text{(paramagnetic phase)};$$
        \item for $\rho = \rho_c$, the minimisers are of either form.
    \end{itemize}
The first derivative of the ground state energy $\rho\mapsto E^{\rm HF}(\rho)$ has a jump at $\rho=\rho_c$. 
    
\medskip

\noindent \textbf{(Second-order phase transition).} In dimension $d\geq3$, if $\min(2,d)<s<d$, then there is a smooth transition between a ferromagnetic and a paramagnetic phase, occurring between two densities
    \[
\rho_{c, {\rm min}}= \frac12 \left( \dfrac{\kappa(d)}{\lambda(d,s)} \dfrac{2(d+2)}{s(d+s)} \right)^{\frac{d}{s-2}}
\qquad\text{and}\qquad 
\rho_{c, {\rm max}}=
\left( \dfrac{\kappa(d)}{\lambda(d,s)} \dfrac{d+2}{d+s} \right)^{\frac{d}{s-2}}.
    \]
More precisely, 
    \begin{itemize}
        \item for $0 < \rho \leq \rho_{c, {\rm min}}$, the minimiser of $E^\HF(\rho)$ is unique, given by
         $$\gamma_{\rm para}(\bk)=\1\big(k^2\leq c_{\rm TF}(\rho/2)^{2/d}\big)\, \bbI_2 \qquad \text{(paramagnetic phase)};$$
        \item for $\rho_{c, {\rm min}} < \rho<   \rho_{c, {\rm max}}$, the minimisers of $\cE^{\rm HF}(\rho)$ are all of the form 
        $$\gamma_{\rm mixed}(\bk)=U\begin{pmatrix}
        \1\big(k^2\leq c_{\rm TF} [t_\rho \rho]^{2/d}\big)&0\\
        0&\1\big(k^2\leq c_{\rm TF}[(1-t_\rho)\rho]^{2/d}\big)
        \end{pmatrix}U^*$$
        with $U\in \SU(2)$, for a unique $t_\rho\in (0,1/2)$;
        \item for $\rho\geq \rho_{c, {\rm max}} $, the minimisers of $\cE^{\rm HF}(\rho)$ are all of the form 
    $$\gamma_{\rm ferro}(\bk)=\1\big(k^2\leq c_{\rm TF}\rho^{2/d}\big)\;|\nu\rangle\langle\nu| \qquad \text{(pure ferromagnetic phase)}$$ 
        with $\nu$ any normalised vector in $\C^2$.
    \end{itemize}
\noindent \textbf{(The case $s = 2$).} If $d \ge 3$ and $s = 2$, then the system is paramagnetic for all $\rho>0$ if $(\kappa(d) - \lambda(d,2)) > 0$, and is pure ferromagnetic for all $\rho > 0$ if $(\kappa(d) - \lambda(d,2)) < 0$.
\end{theorem}

In the three-dimensional Coulomb case, we have $c_{\rm D}(3,1) = 4 \pi^2$, see~\cite[Equation 6.1.21]{ParYan-94}, hence
$$
\kappa(3)=\frac{3^{5/3}\pi^{4/3}}{2^{1/3}5},\qquad \lambda(3,1)=\frac{3^{4/3}}{2^{5/3} \pi^{1/3}}.
$$
This gives the value $\rho_c := 125(24 \pi^5)^{-1} \left( 1+2^{1/3}\right)^{-3}$ claimed in~\eqref{eq:numericalValueRhoc_pre}.

\medskip

According to Theorem~\ref{thm:spin_as_nospin}, the proof of Theorem~\ref{thm:T_0} is reduced to studying the function $P_\rho : [0, 1/2] \to \R$ defined by
\begin{align} \label{eq:P_rho}
P_\rho(t) & := E_\nospin^{\rm HF}(t\rho)+E_\nospin^{\rm HF}((1-t)\rho)\\
&=\kappa(d) \rho^{1+\frac{2}{d}}\Big(t^{1+\frac{2}{d}}+(1-t)^{1+\frac{2}{d}}\Big)
- \lambda(d,s) \rho^{1+\frac{s}{d}}\Big(t^{1+\frac{s}{d}}+(1-t)^{1+\frac{s}{d}}\Big). \nonumber
\end{align}
We set
\begin{equation*}
\lambda := \dfrac{\lambda(d,s)}{\kappa(d) \rho^{\frac{2-s}{d}}}, \qquad q = \frac{d+2}{d} \quad \text{and} \quad p = \frac{d+s}{d},
\end{equation*}
Note that $\lambda = \lambda(\rho)$ is decreasing when $s < 2$ and is increasing when $s > 2$, and that $\frac{2-s}{d} = q - p$. Theorem~\ref{thm:T_0} is a direct consequence of the following lemma.

\begin{lemma} \label{lem:technical} Define the function
\[
    f_\lambda : x \in [0, 1/2] \mapsto x^q + (1 - x)^q - \lambda (x^p + (1 - x)^p).
\]

\medskip

 \noindent    {\bf (Case $s < 2$)} For all $p \in (1,2)$, all $q \in (1, 3)$ with $p < q$ and all $\lambda \ge 0$, the minimum of $f_\lambda$ on $[0, 1/2]$ is $x = 0$ if $\lambda > \lambda_c$ or $x = 1/2$ if $\lambda < \lambda_c$, where
 $
    \lambda_c := \tfrac{1 - 2^{1 - q}}{1 - 2^{1 -p}}.
 $
    
    \medskip
    
 \noindent     {\bf (Case $s > 2$)} For all $1 < q < p < 2$, the minimum of $f_\lambda$ on $[0, 1/2]$ is $\frac12$ if $\lambda \le \lambda_{\rm min} := \frac{q(q-1)}{p(p-1)}2^{p-q}$, and is $0$ if $\lambda \ge \lambda_{\rm max} := \frac{q}{p}$. If $\lambda \in (\lambda_{\rm min}, \lambda_{\rm max})$, then the minimum is strictly between $0$ and $\frac12$, and varies smoothly between $\frac12$ and $0$ as $\lambda$ varies between $\lambda_{\rm min}$ and $\lambda_{\rm max}$. \\
 \noindent {\bf (Case $s = 2$)} If $p = q > 1$, then the minimum of $f_\lambda$ on $[0, 1/2]$ is $\frac12$ if $\lambda < 1$, and is $0$ if $\lambda > 1$.
\end{lemma}

\begin{proof}
   For all $x \in (0, 1/2)$, we have $f_\lambda'(x) = 0$ if and only if
    \[
        q \left[ x^{q-1} - \left(1 - x \right)^{q-1} \right] = p \lambda \left[ x^{p-1} - \left( 1-x \right)^{p-1} \right],\]
or
\[        \lambda = \lambda(x) = \frac{q}{p} \dfrac{ x^{q-1} - \left( 1-x \right)^{q-1} }{ x^{p-1} - (1-x)^{p-1}}.
    \]
   Let us study the map $x \mapsto \lambda(x)$, and prove that it is one-to-one. We have 
    \[
        \lambda'(x) = \frac{p}{q} \left( \dfrac{(1-x)^{q-1} - x^{q-1} }{ (1-x)^{p-1} - x^{p-1}}  \right) \left( \cF_x(q) - \cF_x(p) \right)
    \]
   where we set
    \[
        \cF_x(p) := (p-1) \dfrac{x^{p-2} + (1-x)^{p-2}}{x^{p-1} - (1-x)^{p-1}}.
    \]
One sees that for all $x \in (0, 1/2)$, there exists $p_x \in (2,3)$ such that $\cF_x$ is strictly increasing on $(1, p_x)$, and strictly decreasing on $(p_x, 3)$. In addition, $\cF_x(2) = \cF_x(3) = (x-\frac12)^{-1}$. 
      
   In the case where $p \in (1,2)$ while $q \in (1,3)$ with $p < q$, this implies $\cF_x(p) < \cF_x(q)$. In this case, the map $\lambda(x)$ is strictly increasing. The inverse map $\lambda \mapsto x_\lambda$ is therefore also strictly increasing, and the point $x_\lambda$ is the only critical point of $f_\lambda$ on $(0, 1/2)$. It remains to prove that it is a local maximum. We have $\partial_\lambda \partial_x f_\lambda (x) = \lambda (p-1)((1-x)^{p-1} - x^{p-1}) \ge 0$ for all $x \in (0, 1/2)$. As a result, the map $\lambda \mapsto f_\lambda'(x)$ is strictly decreasing, and vanishes only at $x = x_\lambda$. This implies that $f_\lambda'(x)$ is positive for $x < x_\lambda$, while $f_\lambda'(x)$ for $x > x_\lambda$. In other words, $f_\lambda$ is increasing on $[0, x_\lambda]$ and decreasing on $[x_\lambda, \frac12]$, hence its minimum can only be $0$ or $1/2$. The proof follows.
 
   In the case $1 \le q < p < 2$, the map $\lambda \mapsto \lambda(x)$ is decreasing, with $\lambda(0) = \frac{q}{p}$ and $\lim_{x \to \frac12} \lambda(x) = \frac{q(q-1)}{p(p-1)}2^{p-q}$. Reasoning as before, we see that $f_\lambda$ is decreasing if $\lambda \le \lambda_{\rm min}$, is decreasing then increasing if $\lambda_{\rm min} < \lambda < \lambda_{\rm max}$, and increasing if $\lambda \ge \lambda_{\rm max}$.
\end{proof}

\begin{figure}[h]
    \centering
\includegraphics[width=12cm]{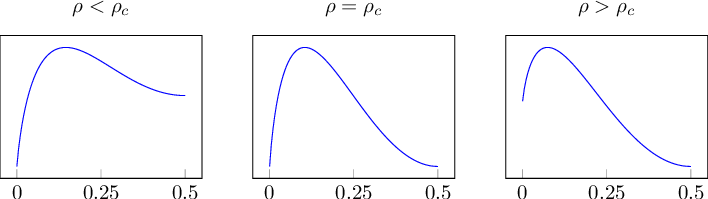}
    \caption{The map $t \mapsto P_\rho(t)$ for different values of $t$ in the case $d= 3$ and $s = 1$ (sharp transition).}
    \label{fig:sharpTransition}
\end{figure}

In order to visualise these phase transitions, we plot the function $P_\rho$ defined in~\eqref{eq:P_rho} for different values of $\rho$. In Figure~\ref{fig:sharpTransition}, we took the three dimensional Coulomb case $d = 3$ and $s = 1$, which corresponds to the sharp transition case. We clearly see that the minimum of $P_\rho$ jumps from $t = 0$ to $t = \frac12$ at the critical density $\rho_c$.

\begin{figure}[h]
    \centering
\includegraphics[width=12cm]{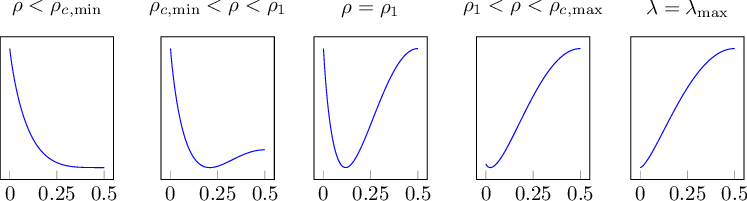}
    \caption{The map $t \mapsto P_\rho(t)$ for different values of $t$ in the case $d= 3$ and $s = 5/2$ (smooth transition). Here, $\rho_1$ is a density in $( \rho_{c, {\rm min}},  \rho_{c, {\rm max}})$.}
    \label{fig:smoothTransition}
\end{figure}

In Figure~\ref{fig:smoothTransition}, we took the values $d = 3$ and $s = 5/2$, which corresponds to the smooth transition case. 

\subsection{Non trivial phase transitions for other interactions}

In this section, we exhibit an example of a repulsive interaction for which complex phase transitions occur as the density $\rho$ increases, in the case $T = 0$. Our example is a combination of Riesz interactions, of the form
\[
    w(\bx) := \dfrac{\alpha_1}{| \bx |^{s_1}} + \dfrac{\alpha_2}{| \bx |^{s_2}} \quad \text{with} \quad  \alpha_{1} > 0, \ \alpha_2 > 0 \quad \text{and} \quad 0 < s_1, s_2 < d.
\]

\begin{figure}[H]
    \centering
\includegraphics[width=9cm]{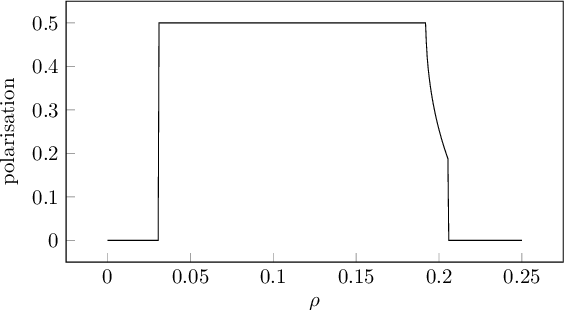}
    \caption{The arg-minimum of $\widetilde{P}_\rho$ as a function of $\rho$.}
    \label{fig:doubleTransition_polarisation}
\end{figure}

Following the lines of the previous section, we are lead to study the minimum of $\widetilde{P_\rho} : [0, \frac12] \to \R$ defined by
\begin{align*}
\widetilde{P_\rho}(t) := & \rho^{1+\frac{2}{d}}\Big(t^{1+\frac{2}{d}}+(1-t)^{1+\frac{2}{d}}\Big) 
- \sum_{i=1,2}  \lambda_i  \rho^{1+\frac{s_i}{d}}\Big(t^{1+\frac{s_i}{d}}+(1-t)^{1+\frac{s_i}{d}}\Big),
\end{align*}
where we set $\lambda_i := \alpha_i  \lambda(d,s_i) / \kappa(d)$. In Figure~\ref{fig:doubleTransition_polarisation}, we plot the arg-minimum $t \in [0, 1/2]$ of $\widetilde{P}_\rho$ as a function of $\rho$. We took the values $d = 3$, $\lambda_1 = 1/2$, $\lambda_2 = 1$, $s_1 = 1/5$ and $s_2 = 14/5$. In this figure, we see that the system is ferromagnetic at low density. Then, the system undergoes a first-order transition to a paramagnetic state around $\rho \sim 0.04$. The system then stays paramagnetic until $\rho \sim 0.18$, where a second-order phase transition to the ferromagnetic state happens. Finally, around $\rho \sim 0.2$, the system becomes suddenly ferromagnetic again.

\begin{figure}[h]
    \centering
\includegraphics[width=10cm]{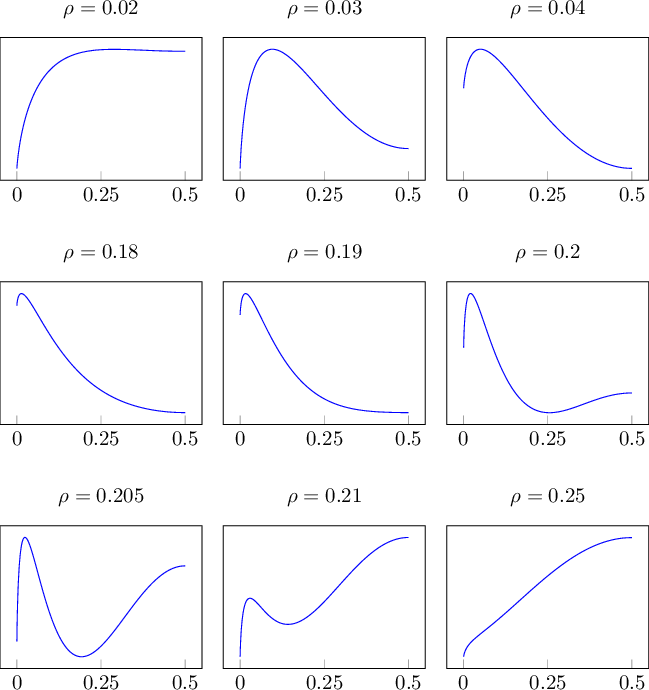}
    \caption{The function $\widetilde{P}_\rho$ for different values of $\rho$.}
    \label{fig:doubleTransition}
\end{figure}

To better illustrate the previous behaviour, we plot in Figure~\ref{fig:doubleTransition} the function $\widetilde{P}_\rho$ for different values of $\rho$. In these figures, we see how the minimum of $\widetilde{P}_\rho$ can change suddenly or smoothly depending on the density.

\section{Uniqueness at $T>0$: proof of Theorems~\ref{th:uniqueness_positive_T} and~\ref{th:uniqueness_paramagnetism_positive_T}}
    \label{sec:proof_thm_positive_T}

In this section, we first prove that the no-spin functional $E^{\rm HF}_\nospin(\rho, T)$ has a unique critical point in the region $\Omega$ defined in Theorem~\ref{th:uniqueness_positive_T}. By Lemma~\ref{lem:well-posed}, we already know that there exist minimisers for $E^{\rm HF}_\nospin(\rho, T)$. We are only interested in the uniqueness here. At the very end, we provide the proof of Theorem~\ref{th:uniqueness_paramagnetism_positive_T}.


Any critical point of $\cE^\HF_\nospin(\cdot, T)$ on the set of states with density $\rho$ satisfies the Euler-Lagrange equation
\begin{equation}
\begin{cases}
\dps g=\frac{1}{\re^{\beta(k^2/2-\mu-g\ast w)}+1},\\[0.3cm]
\dps\frac{1}{(2\pi)^d}\int_{\R^d}g(k)\,{\rm d}k=\rho,
\end{cases}
\label{eq:EulerLagrange_g_proof}
\end{equation}
with $g \in L^1(\R^d) \cap L^\infty(\R^d)$ and $\mu \in \R$. In terms of the potential
$V:=g\ast w \in L^\infty(\R^d)$, this can be rewritten
\begin{equation}
\begin{cases}
\dps V=w\ast \frac{1}{\re^{\beta(k^2/2-\mu-V)}+1},\\[0.3cm]
\dps\frac{1}{(2\pi)^d}\int_{\R^d}\frac{{\rm d}k}{\re^{\beta(k^2/2-\mu-V)}+1}=\rho.
\end{cases}
\label{eq:EulerLagrange_V_proof}
\end{equation}
We prove in this section that Equations~\eqref{eq:EulerLagrange_g_proof}-\eqref{eq:EulerLagrange_V_proof} have a unique solution whenever $(\rho, T) \in \Omega$. In the sequel, we work mainly with~\eqref{eq:EulerLagrange_V_proof}, since $V$ lies in the simple space $L^\infty(\R^d)$. Of course, there is a one-to-one correspondence between the solutions of~\eqref{eq:EulerLagrange_g_proof} and~\eqref{eq:EulerLagrange_V_proof} with $V = g \ast w$ and $g = (\re^{\beta(k^2/2 - \mu - V)} + 1)^{-1}$. 

In our proof we write 
$$\Omega=\Omega_1\cup\Omega_2$$ 
and use different arguments in each of the two sub-domains. In the sub-critical case~\eqref{eq:bound_hat_w} we define the two domains by 
\begin{equation} \label{eq:def:Omega1}
    \Omega_1 := \left\{ (\rho, T) \in \R_+^* \times \R_+^*, \quad \left( C_1 \rho^{s_1/d} + C_2 \rho^{s_2/d} \right) < T  \right\}
\end{equation}
with 
\begin{equation} \label{eq:def:Ci}
    C_i := \kappa_i \frac{| \SS^{d-1} |}{s_i} c_{\rm TF}^{s_i/2}, \quad i  = \{1,2\}
\end{equation}
and 
\begin{equation} \label{eq:def:Omega2}
    \Omega_2 := \left\{  (\rho, T) \in \R_+^* \times \R_+^*, \quad \left( C_1 \rho^{s_1/d} + C_2 \rho^{s_2/d} \right) > \frac{T}2, \quad \rho > C \right\},
\end{equation}
for some large enough $C$. Similarly, in the case $w(k)=\kappa/|k|^{d-1}$ we introduce the two sets 
\begin{equation} \label{eq:def:Omega1_critical}
    \Omega_1 := \left\{ (\rho, T) \in \R_+^* \times \R_+^*, \quad C_1 \rho^{1/d} < T  \right\},
\end{equation}
and 
\begin{equation} \label{eq:def:Omega2_critical}
    \Omega_2 := \left\{  (\rho, T) \in \R_+^* \times \R_+^*, \quad C_1\rho^{1/d} > \frac{T}2, \quad T \re^{ \alpha \rho^{\frac{1}{d}}} >  C \right\}
\end{equation}
for some $\alpha,C>0$ to be determined later. 

In the region $\Omega_1$, we prove below that any solution $(V, \mu)$ of~\eqref{eq:EulerLagrange_V_proof} is non-degenerate, hence gives rise to a smooth branch of solutions in a neighbourhood, by the implicit function theorem. Propagating this information, we get a branch over the whole connected set $\Omega_1$.  In the region $\Omega_2$, we can only prove that any {\bf radial-decreasing} solution $(V, \mu)$ of~\eqref{eq:EulerLagrange_g_proof} is non-degenerate, hence gives rise to a smooth branch of solutions in a neighbourhood. The difficulty here is that although the solution obviously stays radial, it is not obvious that it stays decreasing, hence we cannot easily extend the branch to the whole $\Omega_2$ and finally to $\Omega=\Omega_1\cup\Omega_2$. In order to do so, we work with two special solutions $V_{\rm min}$ and $V_{\rm max}$ that we know are always radial decreasing. We prove that the local branch constructed at a $V_{\rm min/max}$ must stay equal to the solution $V_{\rm min/max}$ in a neighbourhood, hence in particular must stay radial-decreasing. These solutions therefore give rise to branches over the whole set $\Omega$. Finally, we prove that $V_{\rm min}=V_{\rm max}$ in $\Omega_1$, hence we get $V_{\rm min}=V_{\rm max}$ everywhere. 

We remark that it is possible to show that all the solutions $V$ of the fixed point equation~\eqref{eq:EulerLagrange_V_proof} are necessarily radial decreasing, at least under some mild regularity assumption on $w$. This is explained in Appendix~\ref{app:moving_planes}. This additional  information does not simplify our proof, however.

\subsection{Non degeneracy of the linearised operator}
In this section, we investigate the non-degeneracy of solutions. Discarding for the moment the density constraint, we see that the linearised operator for the first equation in~\eqref{eq:EulerLagrange_V_proof} is equal to 
$$\cK_g(v) := v-\beta w\ast \Big(g(1-g)v\Big).$$
with of course $g := ( \re^{\beta(k^2/2 - V - \mu)} +1 )^{-1}$.
Let us introduce the operator 
\begin{equation}  \label{eq:def:Ag}
\boxed{A_g(v):= \beta  w\ast \Big(g(1-g)v\Big)}
\end{equation}
so that $\cK_g=1- A_g$. The following proposition is a key tool in our analysis.

\begin{proposition}[Non-degeneracy in $\Omega$] \label{prop:beta_rho<1}
    Assume $w \le \kappa_1 | \cdot |^{s_1 - d} + \kappa_2 | \cdot |^{s_2 - d}$ with $\kappa_1, \kappa_2 \ge 0$ and $1 < s_1 \le s_2 < \min(d,2)$ and define $\Omega_1$ and $\Omega_2$ by~\eqref{eq:def:Omega1} and~\eqref{eq:def:Omega2}, respectively. Then there exists $C > 0$ such that if $(V, \mu)$ is a solution to~\eqref{eq:EulerLagrange_V_proof} with 
$$(\rho,T) \in \Omega_1$$
or
$$(\rho,T) \in \Omega_2\quad\text{and $V$ is \textbf{radial-decreasing}},$$
then  we have 
$$\| A_g \|_{L^\infty \to L^\infty} < 1.$$
The same results hold for $w(k)=\kappa |k|^{1-d}$ and $d\geq2$, with this time $\Omega_1$ and $\Omega_2$ given by~\eqref{eq:def:Omega1_critical} and~\eqref{eq:def:Omega2_critical}, respectively.
\end{proposition}

We postpone the proof of Proposition~\ref{prop:beta_rho<1} until Section~\ref{ssec:proofAg}. Under the conditions of Proposition~\eqref{prop:beta_rho<1}, we have $\| A_g \|_{L^\infty \to L^\infty} <1$, and the operator $\cK_g$ is invertible on $L^\ii(\R^d)$ with bounded inverse, given by
\begin{equation} \label{eq:inverse_Kg}
(\cK_g)^{-1}=\sum_{n\geq0}(A_g)^n.
\end{equation}
Because of the density constraint in~\eqref{eq:EulerLagrange_V_proof}, we have to work in $L^\ii(\R^d)\times\R$ and the total linearised operator is, for $v \in L^\infty(\R^d)$ and $\delta \in \R$,
$$\cK_{\rm tot}\begin{pmatrix}
v\\ \delta
\end{pmatrix}
:=\begin{pmatrix}
\cK_g(v)-\delta \left[ \beta w\ast g(1-g) \right]\\
\dps\frac{1}{(2\pi)^d}\int_{\R^d}g(1-g)(v+\delta)
\end{pmatrix}.$$
In order to prove that this operator is invertible, we need to solve the system of equations
$$\begin{cases}
\cK_g(v)- \delta \left[\beta w\ast g(1-g) \right] =f\\
\dps\frac{1}{(2\pi)^d}\int_{\R^d}g(1-g)(v+\delta)=\eta
\end{cases}$$
for any given $(f,\eta)\in L^\ii(\R^d)\times\R$. Similarly as for the Schur formula, we can solve the first equation using the invertibility of $\cK_g$ and insert it in the second. We find that $\delta$ has to solve the equation 
\begin{equation} \label{eq:equation_delta}
\int_{\R^d}g(1-g)(\cK_g)^{-1}[f]+\delta\int_{\R^d}\Big(g(1-g)+\beta(\cK_g)^{-1}\left[ w\ast g(1-g) \right]\Big)=(2\pi)^d\eta.
\end{equation}

Since $A_g$ has a positive kernel, so has $(K_g)^{-1}$ by~\eqref{eq:inverse_Kg}. So $(\cK_g)^{-1}\left[ w\ast g(1-g) \right]$ is a positive function, and the coefficient of $\delta$ is at least equal to $\int_{\R^d}g(1-g) > 0$. This proves that~\eqref{eq:equation_delta} has a unique solution $\delta \in \R$. We then find $v$ with
\[
    v = (\cK_g)^{-1} \left(f + \delta [\beta w *g(1-g)]  \right).
\]
This shows that $\cK_{\rm tot}$ is invertible on $L^\ii(\R^d)\times\R$ with bounded inverse whenever $\cK$ is itself invertible.

We obtain the following result.

\begin{corollary}[Construction of branches of solutions]\label{cor:branches}
Under the assumptions of Proposition~\ref{prop:beta_rho<1}, any solution $(V,\mu)$ to~\eqref{eq:EulerLagrange_V_proof} with 
$(\rho,T) \in \Omega_1$ gives rise to a unique real-analytic branch of solution over the whole domain $\Omega_1$. Similarly, any solution $(V,\mu)$ to~\eqref{eq:EulerLagrange_V_proof} with $(\rho,T) \in \Omega_2$ and $V$ radial-decreasing, gives rise to a unique real-analytic branch of solution in a neighbourhood of $(\rho,T)$ in $\Omega_2$.

Finally, on any such branch of solutions, we have
\begin{equation}
\frac\partial{\partial\rho} E^{\rm HF}_\nospin(\rho, T)=\mu(\rho, T),\qquad  \frac\partial{\partial\rho} \mu(\rho, T)=T\frac{(2\pi)^d}{\int_{\R^d}g(1-g)}>0.
 \label{eq:derivatives_rho}
\end{equation}
In particular, $\rho\mapsto E^{\rm HF}_\nospin(\rho, T)$ is a convex function of $\rho$ in the corresponding region. 
\end{corollary}

\begin{proof}
The result follows from the implicit function theorem and the non-degeneracy discussed in this section. The derivative of the energy is found after differentiating and using the implicit function theorem. The derivative of $\mu$ follows from differentiating the second equation in~\eqref{eq:EulerLagrange_V_proof}.
\end{proof}

As mentioned earlier, the branch in $\Omega_2$ cannot easily be extended, since the decreasing property might be lost. In order to see that this is not the case, we now consider two particular solutions which we can extend to the whole set $\Omega$. 

\subsection{Construction of $V_{\rm min}$ and $V_{\rm max}$}

\begin{proposition}[The minimal and maximal solutions]\label{prop:V_min_max}
Under the assumptions of Proposition~\ref{prop:beta_rho<1}, there exist two real-analytic branches of solutions of~\eqref{eq:EulerLagrange_V_proof} in the whole set $\Omega$, denoted by $(V_{\rm min}(\rho,T),\mu_{\rm min}(\rho,T))$ and $(V_{\rm max}(\rho,T),\mu_{\rm max}(\rho,T))$, which satisfy the properties that, for any $(\rho,T)\in\Omega$,
\begin{itemize}
 \item $V_{\rm min}(\rho,T)$ and $V_{\rm max}(\rho,T)$ are radial-decreasing;
 \item for any other solution $(V,\mu)$ of~\eqref{eq:EulerLagrange_V_proof} for $(\rho', T)$ with $\mu=\mu_{\rm min}(\rho,T)$, we have
 $$V\geq V_{\rm min}(\rho,T)$$
 pointwise. In particular, $\rho' \ge \rho$;
 \item for any other solution $(V,\mu)$ of~\eqref{eq:EulerLagrange_V_proof} for $(\rho', T)$ with $\mu=\mu_{\rm max}(\rho,T)$, we have
 $$V\leq V_{\rm max}(\rho,T)$$
 pointwise. In particular, $\rho' \le \rho$.
\end{itemize}
\end{proposition}

In the next section we prove that the two branches are actually equal. The rest of the section is devoted to the proof of Proposition~\ref{prop:V_min_max}. 

For any $T>0$ and $\mu\in\R$, we study the fixed point equation
\begin{equation}
V = \cV_{\mu, T} (V), \quad \text{where} \quad \cV_{\mu, T}(U):= w * \frac{1}{\re^{\beta(k^2/2-\mu-U)}+1}.
\label{eq:Hammerstein_map}
\end{equation}
The map $\cV_{T, \mu}$ is order-preserving, in the sense that if $U_1\leq U_2$, then $\cV_{T,\mu}(U_1)\leq  \cV_{T,\mu}(U_2)$ pointwise. In addition, if $U$ is radial non-increasing, then $\cV_{T,\mu}(U)$ is radial decreasing. 

\begin{lemma} \label{lem:critical_points_are_bounded}
    There is $C > 0$ large enough such that, for all $V \in L^\infty(\R^d)$, all $T > 0$ and all $\mu \in \R$ such that $V = \cV_{\mu, T} (V)$, then 
    \begin{equation}
\| V \|_{L^\infty} \le C \left(1 + | \mu | + \beta^{-s_1/2}+\beta^{-s_2/2} \right).
\label{eq:estim_V_infty}
    \end{equation}
\end{lemma}

The lemma applies to the critical case $s_1=1$ and $\kappa_2 = 0$ in dimension $d\ge 2$.

\begin{proof}[Proof of Lemma~\ref{lem:critical_points_are_bounded}]
    First, since $V$ is a fixed point of~\eqref{eq:Hammerstein_map}, we have $V > 0$. For shortness we denote
    $$a := \| V \|_{L^\infty(\R^d)} > 0.$$
    If $a \le | \mu |$, our lemma is proved. We now assume that $a \ge |\mu|$.
    Since $0 \le V \le a$, we have $\cV_{\mu, T}(V) \le \cV_{\mu, T}(a)$, hence $a \le \| \cV_{\mu, T}(a) \|_{L^\infty(\R^d)}$. The function $\cV_{T, \mu}(a)$ is radial decreasing, so
    \[
    \| \cV_{T, \mu} (a) \|_{L^\infty} = \cV_{T, \mu} (a) (0)
        = \int_{\R^d}  \dfrac{w(k)\,\rd k}{\re^{\beta (k^2/2 - \mu - a)} + 1}
        \le \int_{\R^d} \dfrac{w(k)\,\rd k}{\re^{\beta (k^2/2 -2a)} + 1},
    \]
    where we used the fact that $\mu \le | \mu | \le a$ for the last inequality. For all $1 \le s < \min(d,2)$, we have by scaling
    \begin{align*} 
    \int_{\R^d} \dfrac{1}{| k |^{d-s}} \dfrac{\rd k}{\re^{\beta (k^2/2 - 2a)}+1} &
    = (2a)^{s/2} \int_{\R^d} \dfrac{1}{| y |^{d-s}} \dfrac{\rd y}{\re^{\beta 2a (y^2/2 - 1)} + 1}.
    \end{align*}
    Let us study the function
    \begin{equation} \label{eq:def:J}
        J(x) := \int_{\R^d} \dfrac{1}{| y |^{d-s}} \dfrac{\rd y}{\re^{x (y^2/2 - 1)} + 1}.
    \end{equation}
    For all $x > 0$, $J(x) > 0$, and when $x \to \infty$, $J(x)$ goes to the finite value $J(\infty) = \int_{\R^d} | y |^{s-d} \1 (y^2 \le 2) \rd y$. Also, we have
    \[
        J(x) \le  \int_{\R^d} \dfrac{1}{| y |^{d-s}} \re^{-x (y^2/2 - 1)} \rd y = \dfrac{\re^{x}}{ x^{s/2} } \left( \int_{\R^d} \dfrac{1}{| y |^{d-s}} \re^{- \frac{y^2}{2}} \rd y \right).
    \]
    We deduce that there $0 < c_1(s) \le c_2(s)$ such that $c_1(s) \le J(x) \le c_2(s) \max(1, x^{-s/2})$. This finally gives
    \[
        a \le \| \cV_{T, \mu}(a) \|_{L^\infty} \le \sum_{i=1}^2 \kappa_i c_2(s_i) \max \left( (2a)^{s_i/2}, \dfrac{1}{\beta^{s_i/2}} \right)
    \]
with the convention that $s_1=1$ and $\kappa_2=0$ in the critical case. When $a$ is large enough and since $s_i < 2$, this gives 
   \[
        a \le C+\sum_{i=1}^2  \dfrac{\kappa_i c_2(s_i)}{\beta^{s_i/2}},
    \]
which concludes the proof.
\end{proof}

We now prove that there is a unique solution smaller than any other solution, and another one greater than any other solution.

\begin{lemma}[Existence of $V_{\rm min}$ and $V_{\rm max}$]\label{lem:existence_V_min_max}
For any fixed $T>0$ and $\mu\in\R$, there exist two unique radial-decreasing functions $V_{\rm min}(\mu,T)$ and $V_{\rm max}(\mu,T)$ solutions of~\eqref{eq:Hammerstein_map} satisfying that  $V_{\rm min}(\mu,T)\leq V\leq V_{\rm max}(\mu,T)$ for any other solution $V$ of~\eqref{eq:Hammerstein_map}. We denote by
$$g_{\rm min/max}(\mu,T):=\frac{1}{e^{\beta(k^2/2-\mu-V_{\rm min/max}(\mu,T))}+1}$$
the corresponding states and by
$$\rho_{\rm min/max}(\mu,T):=\frac{1}{(2\pi)^d}\int_{\R^d}g_{\rm min/max}(\mu,T)$$
the corresponding densities. 
\end{lemma}

In the terminology of Hammerstein integral equations~\cite{BroGup-69}, $V_{\rm min}(\mu, T)$ and $V_{\rm max}(\mu, T)$ are respectively the minimal and maximal fixed points of the increasing map $\cV_{\mu, T}$. The two families of solutions $V_{\rm min/max}$ are parametrised by $\mu$ and $T$. These are not necessarily continuous and the corresponding $\rho_{\rm min/max}$ is not necessarily one-to-one. Later we will restrict our attention to the solutions lying in $\Omega$, which will be well behaved. 

\begin{proof}[Proof of Lemma~\ref{lem:existence_V_min_max}]
The same computation as in the proof of Lemma~\ref{lem:critical_points_are_bounded} shows that for $a$ of the order of the right side of~\eqref{eq:estim_V_infty}, (hence depending on $\mu$ and $T$), we have $0 \le \cV_{\mu, T}(a) < a$ pointwise. By induction, we deduce that $\big(\cV_{\mu, T}\big)^n(a)$ is a pointwise decreasing sequence bounded by $0$, hence converges pointwise to a function $V_{\rm max}(\mu, T)$. By continuity of $\cV(\mu, T)$, $V_{\rm max}(\mu, T)$ is a fixed point of $\cV_{\mu, T}$. 

Similarly, we have $0 \le \cV_{\mu, T}(0) \le V_{\rm max}(\mu, T)$, hence the sequence $\big(\cV_{\mu, T}\big)^n(0)$ is pointwise increasing and bounded by $V_{\rm max}(\mu, T)$, hence converges to some $V_{\rm min}(\mu , T)$, which is also a fixed point of $\cV_{\mu, T}$. By construction, both $V_{\rm min}(\mu, T)$ and $V_{\rm max}(\mu, T)$ are radial decreasing.

For any other fixed point $V$ of $\cV_{T, \mu}$, we have $0 \le V \le a$ pointwise by Lemma~\ref{lem:critical_points_are_bounded}, hence, by induction $(\cV_{\mu, T}\big)^n(0) \le V \le (\cV_{\mu, T}\big)^n(a)$, and, after passing to the limit, we find $V_{\rm min}(\mu , T) \le V \le V_{\rm max}(\mu , T)$. 
\end{proof}

In order to conclude the proof of Proposition~\ref{prop:V_min_max}, we also need the following compactness result.

\begin{lemma}[Compactness of critical points] \label{lem:compacity}
    Let $(\mu_n, T_n, V_n)$ be any sequence such that $V_n$ is a fixed point of $\cV_{\mu_n, T_n}(V_n)$. If $\mu_n \to \mu_\infty$ and $T_n \to T_\infty$, then there is a function $V_\infty \in L^\infty(\R^d)$ such that, up to a subsequence, $V_n$ converges strongly to $V_\infty$ in $L^\infty(\R^d)$, and $V_\infty = \cV_{\mu_\infty, T_\infty}(V_\infty)$.
\end{lemma}

\begin{proof}
    Since $\mu_n$ and $T_n$ are bounded, we deduce from Lemma~\ref{lem:critical_points_are_bounded} that $V_n$ is uniformly bounded in $L^\infty(\R^d)$. In particular, $g_n := (\re^{ \beta_n (k^2/2 - \mu_n - V_n)} + 1)^{-1}$ is bounded in $L^1(\R^d) \cap L^\infty(\R^d)$. Up to a subsequence, $g_n$ converges weakly(--$\ast$) to some $g_\infty$ in $L^1\cap L^\ii$. Since $w(x - \cdot)$ belongs to $L^{q_1} + L^{q_2}$ for some $1 < q_1 < q_2 < \infty$, we deduce that $\int w(x - \cdot) g_n$ converges to $\int w(x - \cdot) g_\infty$. In other words, $V_n = w * g_n$ converges pointwise to $V_\infty = w* g_\infty$. 
    
    We now bootstrap the argument, and deduce that $g_n \to g_\infty$ pointwise. By Lemma~\ref{lem:critical_points_are_bounded}, there is $\beta'> 0$ and $\mu' > 0$ such that
    $$0\leq g_n\leq \frac{1}{\re^{\beta_n(k^2/2-\mu_n-\|V_n\|_\ii)}+1}\leq \frac{1}{\re^{\beta'(k^2/2-\mu')}+1}.$$
    Together with the dominated convergence theorem, this proves that $g_n\to g_\infty$ strongly in $L^p$ for all $1\leq p<\ii$. By H\"older's inequality we finally get $V_n=g_n\ast w\to V_\infty=g_\ii\ast w$ strongly in $L^\ii$.
\end{proof}

Now we are able to provide the

\begin{proof}[Proof of Proposition~\ref{prop:V_min_max}]
We construct the two branches as follows. Let $(\rho,T)\in\Omega_1$, the set defined in~\eqref{eq:def:Omega1}, and let $(V,\mu)$ be any solution of the nonlinear equation~\eqref{eq:EulerLagrange_V_proof}, for this value of $\rho$ (for example a minimiser). Then, for this chemical potential $\mu$, the minimal solution satisfies $V_{\rm min}(\mu,T)\leq V$, and therefore $\rho_{\rm min}(\mu,T)\leq \rho$. From the definition of $\Omega_1$, we conclude that $(\rho_{\rm min}(\mu,T),T)\in\Omega_1$. Hence at least one of the minimal solutions lies in $\Omega_1$. Similarly, by considering $(\rho, T) \in \Omega_2$, we prove that at least one of the maximal solutions lies in $\Omega_2$. 

Now we extend the two branches as follows. For shortness we only discuss the minimal solution, since the argument is the same for the maximal solution. We assume that $\rho_0=\rho_{\rm min}(\mu_0,T_0)$ with $(\rho_0,T_0)\in\Omega=\Omega_1\cup\Omega_2$. Since $V_{\rm min}(\mu_0, T_0)$ is radial-decreasing, we may apply Corollary~\ref{cor:branches} to obtain a unique local branch of solutions $(V(\rho,T),\mu(\rho,T))$, by the implicit function theorem. Our goal is to show that this only consists of minimal solutions, that is, $V(\rho,T)=V_{\rm min}(\mu(\rho,T),T)$ for every $(\rho,T)$ in a neighbourhood of the given $(\rho_0,T_0)$. The propagation of the radial-decreasing property allows us to go on with the implicit function theorem and hence, by extension, to obtain a branch over the whole domain $\Omega$. 

So let us assume by contradiction that there exists a sequence $(\rho_n, T_n)\to (\rho_0, T_0)$, and a corresponding sequence $(V_n, \mu_n) := (V(\rho_n, T_n), \mu(\rho_n, T_n))$ that converges to $(V_0, T_0)$ in $L^\infty(\R^d) \times \R$, such that $V_n$ is never the minimal solution for the chemical potential $\mu_n$. Since $V_n$ is a critical point of $\cV_{\mu_n, T_n}$, we have the pointwise estimate.
\begin{equation*}
    V_n > W_n := V_{\rm min}(\mu_n, T_n).
\end{equation*}
By Lemma~\ref{lem:compacity} we have (up to extraction) that $W_n$ converges to some $W_\infty$ which is a fixed point of $\cV_{\mu_0, T_0}$. Since $W_n \le V_n$, we obtain at the limit $W_\infty \le V_{\rm min}(\mu_0, T_0)$. On the other hand, since $W_\infty$ and $V_{\rm min}(\mu_0, T_0)$ are both fixed points of $\cV_{\mu_0, T_0}$, we have by minimality that $V_{\rm min}(\mu_0, T_0) \le W_\infty$. Hence $W_\infty = V_{\rm min}(\mu_0, T_0)$, and $W_n$ converges to $V_{\rm min}(\mu_0, T_0)$ strongly in $L^\ii(\R^d)$. Finally, by uniqueness of the branch in the neighbourhood, we must have $W_n = V_n$ for $n$ large enough, which is the desired contradiction. We deduce, as we wanted, that we can define a branch of minimal solutions, and a branch of maximal solution, on the whole set $\Omega$.
\end{proof}

\subsection{Equality of the minimal and maximal solutions}

In the previous section, we have constructed two smooth branches of solution $(V_{\rm min}(\rho, T), \mu_{\rm min}(\rho, T) )$ and $(V_{\rm max}(\rho, T), \mu_{\rm max}(\rho, T) )$ on the whole set $\Omega$. We now prove that these two branches coincide. Thanks to the implicit function theorem, it is enough to prove that they coincide at a single (and well-chosen) point $(\rho_0, T_0) \in \Omega$.

\begin{proposition} \label{prop:same_mu}
    For all $T > 0$, there is $0 < \rho_1 \le \rho_2$ with $(\rho_1, T) \in \Omega_1$ and $(\rho_2, T) \in \Omega_1$ such that
    $\mu_{\rm min}(\rho_1, T) = \mu_{\rm max}(\rho_2, T)$.
\end{proposition}

\begin{proof}
    Let $\rho_2  > 0$ be such that $(\rho_2, T) \in \Omega_1$. We assume first that $\mu_{\rm min}(\rho_2, T) \ge \mu_{\rm max}(\rho_2, T )$. Since $(0, \rho_2] \times \{T\} \in \Omega_1$, we deduce from Corollary~\ref{cor:branches} that the map $\rho \mapsto \mu_{\rm min}(\rho, T)$ is continuous and increasing on $(0, \rho_2]$. On the other hand, we have
    \[
        g_{\rm min} \ge \dfrac{1}{1 + \re^{\beta(k^2/2 - \mu_{\rm min})}}.
    \]
    Integrating, this shows that $\rho \ge I_\beta(\mu_{\rm min}(\rho, T))$, where
    \begin{equation} \label{eq:def:Ibeta}
        I_\beta(\mu) := \dfrac{1}{(2 \pi)^d} \int_{\R^d} \dfrac{\rd k}{1 + \re^{\beta(k^2/2 - \mu)}}.
    \end{equation}
    The function $I_\beta$ is continuous increasing with $I_\beta(-\infty) = 0$ and $I_\beta(\infty)= \infty$. This proves that $\lim_{\rho \to 0^+} \mu_{\rm min}(\rho, T) = -\infty$, and the proof of Proposition~\ref{prop:same_mu} follows from the mean-value theorem.
    
    In the case where we have $\mu_{\rm max}(\rho_2, T) > \mu_{\rm min}(\rho_2, T )$, we repeat the argument with the map $\rho \mapsto \mu_{\rm max}(\rho, T)$. However, this cannot happen, as we would have for the corresponding solutions,
    \[
        \rho_1 = \dfrac{1}{(2 \pi)^d} \int_{\R^d} g_{\rm max} (\mu, T) \ge \dfrac{1}{(2 \pi)^d} \int_{\R^d} g_{\rm min} (\mu, T) = \rho_2,
    \]
    by minimality of $g_{\rm min}$, which contradicts the fact that $\rho_1 < \rho_2$.
\end{proof}

Let $0 \le \rho_1 \le \rho_2$ be as in Proposition~\ref{prop:same_mu}, so that the corresponding functions $V_{\rm min} := V_{\rm min}(\rho_1, T)$ and $V_{\rm max} := V_{\rm max}(\rho_2, T)$ are two fixed points of the {\em same} map~$\cV_{\mu, T}$ with $\mu := \mu_{\rm min}(\rho_1, T) = \mu_{\rm max}(\rho_2, T)$. We now write
\begin{align*}
       0 \le V_{\rm max} - V_{\rm min} & = 
       w* \left[ \left( 1 + \re^{\beta (\frac{k^2}{2} - V_{\rm max}  - \mu)} \right)^{-1}  - \left( 1 + \re^{\beta (\frac{k^2}{2} - V_{\rm min}  - \mu)} \right)^{-1} \right] \\
       & = w * \left[ \dfrac{1}{ 1 + \re^{\beta (\frac{k^2}{2} - V_{\rm max}  - \mu)}} \dfrac{\re^{\beta (\frac{k^2}{2} - V_{\rm min}  - \mu)}}{ 1 + \re^{\beta (\frac{k^2}{2} - V_{\rm min}  - \mu)} } 
          \left( 1 - \re^{ - \beta (V_{\rm max} - V_{\rm min}) }  \right) \right] \\
        & = w * \left[ g_{\rm max} (1 - g_{\rm min})  \left( 1 - \re^{ - \beta  (V_{\rm max} - V_{\rm min}) } \right) \right] \\
         &\le \beta \| V_{\rm max} - V_{\rm min} \|_{L^\infty} (w * g_{\rm max}),
   \end{align*}
   where we have used that $1 - \re^{ -x} \le x$ for $x \ge 0$ and that $g_{\rm min}\leq 1$. We deduce that
   \begin{equation}
  \left\| V_{\rm max} - V_{\rm min} \right\|_{L^\infty(\R^d)} \leq \beta \norm{w \ast g_{\rm max}}_{L^\ii(\R^d)}  \left\| V_{\rm max} - V_{\rm min} \right\|_{L^\infty(\R^d)}.
   \label{eq:comparison_min_max}
   \end{equation}

   \begin{lemma} \label{lem:estime_g*w}
         Assume $w(k) \le \kappa_1 | k |^{s_1 - d} + \kappa_2 | k |^{s_2 - d}$ with $\kappa_1, \kappa_2 \ge 0$ and $1 \le s_1 \le s_2 < \min(d, 2)$. Then 
        \begin{equation} \label{eq:ineq:g*w}
        \sup\left\{ \norm{w\ast g}_{L^\ii}  , \ 0 \le g \le 1, \ \dfrac{1}{(2 \pi)^d} \int_{\R^d} g = \rho \right\}
        \le  C_1 \rho^{s_1/d} + C_2 \rho^{s_2/d},
        \end{equation}
        where the constants $C_1$ and $C_2$ have been defined in~\eqref{eq:def:Ci}.
   \end{lemma}

The result covers the critical case $s_1=1$ and $\kappa_2=0$, in dimension $d\geq2$. 
   
 \begin{proof}
        We have
        \begin{align*}\norm{w\ast g}_{L^\ii} 
        \le \sup_{\substack{\norm{g}_{L^\ii} \le 1 \\ \norm{g}_{L^1}\leq (2\pi)^d\rho }} \norm{w\ast g}_{L^\ii} 
        \leq \sum_{i=1}^2 \kappa_i \sup_{\substack{\norm{g}_{L^\ii}\leq1\\
                \norm{g}_{L^1}\leq (2\pi)^d\rho}} \left\{ \left\| | \cdot |^{s_i - d} * g \right\|_{L^\infty}\right\}.
        \end{align*}
        By rearrangement inequalities the right side is maximised for $g$ a positive radial decreasing function and by linearity the unique maximiser is $g(k)=\1 \left( k^2 \le C_{\rm TF} \rho^{2/d} \right)$. Hence
        \begin{align}
        \sup_{\substack{\norm{g}_{L^\infty}\leq1\nn\\
                \norm{g}_{L^1}\leq (2\pi)^d \rho}} \norm{|\cdot|^{s-d}\ast g}_{L^\ii}
        &= \sup_{k\in\R^d}\int_{\R^3}\frac{\1 \left( (k')^2 \le c_{\rm TF} \rho^{2/d} \right) {\rm d}k'}{|k-k'|^{d-s}}\nn
        = \dfrac{| \SS^{d-1} |}{s} c_{\rm TF}^{s/2} \rho^{s/d}. \nn
        \end{align}
        This leads to~\eqref{eq:ineq:g*w}.
    \end{proof} 

Applying the lemma we infer, by the definition~\eqref{eq:def:Omega1} of $\Omega_1$ (resp.~\eqref{eq:def:Omega1_critical} in the critical case) that 
$$\beta \norm{w \ast g_{\rm max}}_{L^\ii(\R^d)}<1.$$
Hence we deduce from~\eqref{eq:comparison_min_max} that $V_{\rm min}=V_{\rm max}$. This implies in particular that $\rho_1 = \rho_2$, so the two solutions coincide at the point $(\rho_2, T) \in \Omega$. Altogether, this proves that the two branches $V_{\rm min}(\rho,T)$ and $V_{\rm max}(\rho,T)$ coincide over the whole domain $\Omega$. 

\begin{corollary}[Uniqueness of critical points in $\Omega$]\label{cor:uniqueness_critical_pt}
For any $(\rho,T)\in\Omega$, the function $g\mapsto \cE^{\rm HF}_{\nospin}(g,T)$ has a \emph{unique criticial point} $g_{\rho,T}=g_{\rm min}(\rho,T)=g_{\rm max}(\rho,T)$ of density $\rho$, which is therefore the \emph{unique minimiser} for $E^{\rm HF}_{\nospin}(\rho,T)$.
\end{corollary}

\begin{proof}
Let $(V, \mu)$ be any solution of~\eqref{eq:EulerLagrange_V_proof} for some $(\rho, T) \in \Omega$. We have $V_{\rm min}(\mu, T) \le V \le V_{\rm max}(\mu, T)$. On the other hand, depending whether $(\rho, T) \in \Omega_1$ or $(\rho, T) \in \Omega_2$, we have either $(\rho_{\rm min}(\mu, T), T) \in \Omega_1$ or $(\rho_{\rm max}(\mu, T), T) \in \Omega_2$. In both cases, we deduce that $V_{\rm min}(\mu, T) = V_{\rm max}(\mu, T)$, and finally that $V = V_{\rm min}(\mu, T)$. This concludes the proof. 
\end{proof}

\subsection{Proof of Proposition~\ref{prop:beta_rho<1}: the operator $A_g$ is contracting}
\label{ssec:proofAg}

It remains to prove Proposition~\ref{prop:beta_rho<1}. First, since $A_g$ has a positive kernel, we have
\begin{equation}
\forall f \in L^\infty(\R^d), \quad \| A_g(f) \|_{L^\infty(\R^d)} \le \| f \|_{L^\infty(\R^d)} \| A_g(1) \|_{L^\infty(\R^d)}.
\end{equation}
This shows that $\| A_g \|_{L^\infty \to L^\infty} = \| A_g(1) \|_{L^\infty} = \beta \| w \ast g(1-g) \|_{L^\infty}$. We now give two different estimates in $\Omega_1$ and $\Omega_2$. 

\subsubsection{Estimate in $\Omega_1$}
In the region $\Omega_1$ (defined either by~\eqref{eq:def:Omega1} in the sub-critical case or by~\eqref{eq:def:Omega1_critical} in the critical case), we use that $0 \le g \le 1$, so that $\| A_g \| \le \beta \| w \ast g \|_{L^\infty}$. Together with Lemma~\ref{lem:estime_g*w}, we directly deduce that
\begin{equation*}
\| A_g \|_{L^\infty \to L^\infty} \le \beta \left( C_1 \rho^{s_1/d} + C_2 \rho^{s_2/d} \right).
\end{equation*}
with $s_1=1$ and $\kappa_2 = 0$ in the critical case. 

\subsubsection{Estimate in $\Omega_2$ in the subcritical case~\eqref{eq:bound_hat_w}}
\label{sssec:estimates_in_Omega2}

We now assume that $w(k)\leq \kappa_1|k|^{s_1-d}+\kappa_2|k|^{s_2-d}$ with $1<s_1\leq s_2<2$ and that $\Omega_2$ is given by~\eqref{eq:def:Omega2}. 
In order to estimate the norm of $A_g$ we use ideas from~\cite{GonHaiLew-19} which were based on spectral techniques recently developed in the context of Bardeen-Cooper-Schrieffer theory in~\cite{FraHaiNabSei-07,HaiHamSeiSol-08,HaiSei-08b,FreHaiSei-12,HaiSei-16,HaiLos-17}. We introduce
\begin{equation} \label{eq:def:h}
h(k) := \frac{k^2}{2} - V - \mu,
\end{equation}
so that $g = (1 + \re^{\beta h})^{-1}$. We have 
$$g(1-g) = \frac{1}{2 + 2 \cosh(\beta h)} \le \frac1{4 + \beta^2 h^2},$$ 
where we used the inequality $\cosh(x) \ge 1 + x^2/2$. As in~\cite{GonHaiLew-19}, we note that if two functions $f$ and $g$ are increasing on $\R^+$ with $f(k_*)+g(k_*)=0$, then 
$|f(k)+g(k)|\geq |f(k)-f(k_*)|$. In our case, $k^2/2 - \mu$ and $-V$ are both radial increasing. Let $k_* > 0$ be so that $h(k_*) = 0$ (hence $g(k_*) = 1/2$). We obtain 
\begin{equation} \label{eq:estimate_with_laplacian}
| h(k) | \ge \frac12 |k^2 - k_*^2|,
\end{equation}
 and finally the simple pointwise bound
\[
g(k)(1-g(k)) \le \dfrac{1}{4 + \frac{\beta^2}{4} | k^2 - k_*^2 |^2}.
\]
This proves that 
$$\| A_g \| \le \beta \left\| w * \left(4 + \frac{\beta^2}{4} | k^2 - k_*^2 |^2\right)^{-1} \right\|_{L^\infty}.$$ 
In hyperspherical coordinates, this is
\begin{align}
\| A_g \| & \le \beta  \sup_{\ell \in \R^d} \left( \int_{0}^\infty \dfrac{r^{d-1} \rd r}{4 + \frac{\beta^2}{4} |r^2 - k_*^2 |^2} \int_{\SS^{d-1}} w(\ell - r \omega) \rd \omega \right) \nn \\
& \le  \int_{0}^\infty \dfrac{\beta r^{d-1} \rd r}{4 + \frac{\beta^2}{4} |r^2 - k_*^2 |^2} \sup_{\ell \in \R^d} \left( \int_{\SS^{d-1}} w(\ell - r \omega) \rd \omega \right) \nn \\
& \le \sum_{i=1}^2  \kappa_i\int_{0}^\infty \dfrac{\beta r^{d-1} \rd r}{4 + \frac{\beta^2}{4} |r^2 - k_*^2 |^2} \sup_{\ell \in \R^d} \left( \int_{\SS^{d-1}} \dfrac{ \rd \omega}{| \ell - r \omega |^{d-s_i}} \right)\nn\\
& = \sum_{i=1}^2  \kappa_i\int_{0}^\infty \dfrac{\beta r^{s_i-1} \rd r}{4 + \frac{\beta^2}{4} |r^2 - k_*^2 |^2} \sup_{\ell' \in \R^d} \left( \int_{\SS^{d-1}} \dfrac{ \rd \omega}{| \ell' - \omega |^{d-s_i}} \right). \label{eq:integral_on_sphere}
\end{align}
For $s_i > 1$, the last integral is a bounded function of $\ell'$. Hence we obtain 
$$\| A_g \|\leq C\sum_{i=1}^2  \kappa_i\int_{0}^\infty \dfrac{\beta r^{s_i-1} \rd r}{4 + \frac{\beta^2}{4} |r^2 - k_*^2 |^2}.$$
We have by scaling
\[
\int_{0}^\infty \dfrac{\beta r^{s-1}  \rd r}{4 + \frac{\beta^2}{4} |r^2 - k_*^2 |^2}
= \dfrac{1}{\beta k_*^{4-s}} \int_0^\infty \dfrac{r^{s-1} \rd r}{\frac{4}{\beta^2 k_*^4} + \frac14 | r^2 - 1 |^2}
\le  \dfrac{1}{\beta k_*^{4-s}}  C (1 + \beta k_*^2),
\]
for some large constant $C$, where, for the last inequality, we used the fact that the integrand is integrable at infinity, and has a singularity only at $r = 1$. Altogether, we proved that there is $C >0$ so that 
\[
\| A_g \| \le C \left( \dfrac{1}{\beta k_*^{4 - s_1}} + \dfrac{1}{\beta k_*^{4 - s_2}} + \dfrac{1}{k_*^{2-s_1}} +   \dfrac{1}{k_*^{2-s_2}}   \right) .
\]
We finally use the following technical lemma (proved below), valid in both the sub-critical and critical cases.
\begin{lemma} \label{lem:estimate_k*}
    There exists $0 < c_1 \le c_2$ and $\rho_C>0$  such that, for all $(\rho, T)$ satisfying $\beta^{-1} \le 2 \left(C_1 \rho^{s_1/d} + C_2 \rho^{s_2/d} \right)$ (with as usual $s_1 = s_2 = s$ in the critical case) and $\rho>\rho_C$, we have
    \[
    c_1 \rho^{1/d} \le k_* \le c_2 \rho^{1/d}.
    \]
\end{lemma}
Hence, if $\beta^{-1} \le 2 \left(C_1 \rho^{s_1/d} + C_2 \rho^{s_2/d} \right)$ and $\rho > \rho_C$, we deduce that
\[
\| A_g \|
\le C \left( \rho^{\frac{2(s_1 - 2)}{d}} + \rho^{\frac{2(s_2 - 2)}{d}} + \rho^{\frac{s_1 +s_2 - 4}{d}} +  \rho^{\frac{s_1-2}{d}} + \rho^{\frac{s_2-2}{d}} \right),
\]
which is smaller than $1$ if $\rho_C$ is large enough. This concludes the proof of Proposition~\ref{prop:beta_rho<1} in the sub-critical case.

\begin{proof}[Proof of Lemma~\ref{lem:estimate_k*}]
We define the map $\rho \mapsto \mu_{\rm free}(\rho, T) \in \R$ to be the (unique) solution to $I_\beta(\mu_{\rm free}(\rho, T)) = \rho$, where $I_{\beta}$ was defined in~\eqref{eq:def:Ibeta}. This is the Lagrange multiplier for the free Fermi gas.

We first prove that $c_1 \rho^{2/d} \le \mu_{\rm free}(\rho,T) \le c_2 \rho^{2/d}$ for $\rho$ large enough, independently of $T$. Then we prove a similar inequality for $\mu$ instead of $\mu_{\rm free}$, and finally, we prove the result for $k_*$.

Since $I_\beta$ is increasing, the multiplier $\mu_{\rm free}(\rho, T)$ is positive if and only if 
\[
    \rho > I_\beta(0) = \dfrac{1}{(2 \pi)^d} \dfrac{1}{\beta^{d/2}} \left(  \int_{\R^d} \dfrac{\rd k}{\re^{k^2/2} + 1} \right).
\]
In a region where $\beta^{-1} \le 2 \left(C_1 \rho^{s_1/d} + C_2 \rho^{s_2/d}\right)$, this is the case whenever
\[
    \rho \ge \dfrac{1}{(2 \pi)^d}\left(  \int_{\R^d} \dfrac{\rd k}{\re^{k^2/2} + 1} \right) 2^{d/2} \left(C_1 \rho^{s_1/d} + C_2 \rho^{s_2/d} \right)^{d/2}.
\]
Since $1 \le s_1 \le s_2 < 2$, we deduce that there is $\rho_1 > 0$ such that, for all $(\rho, T)$ such that $\beta \left(C_1 \rho^{s_1/d} + C_2 \rho^{s_2/d}\right) > \frac12$ and $\rho > \rho_1$, then $\mu_{\rm free}(\rho, T) \ge 0$.

By scaling, we have, for all $\mu > 0$, $I_\beta (\mu) = \mu^{d/2} I_{\beta \mu}(1)$. As in the study of the function $J(x)$ defined in~\eqref{eq:def:J}, there is $0 < c \le C$ such that, for all $x> 0$, $c \le I_x(0) \le C \max \{ 1 , x^{-d/2}\}$. This proves that for $\rho \ge \rho_1$,
\[
    c \mu_{\rm free}^{d/2} (\rho, T) \le I_\beta(\mu_{\rm free}(\rho, T)) = \rho \leq C \max \left\{ \mu_{\rm free}^{d/2} (\rho, T), \dfrac{1}{\beta^{d/2}} \right\}.
\]
Again, in a region where $\beta \left(C_1 \rho^{s_1/d} + C_2 \rho^{s_2/d} \right) > 1/2$, $\beta^{-d/2}$ is of lower order compared to $\rho$, hence the maximum is attained for the first member. In other words, there is $\rho_2 > \rho_1$ and $0 < c_1 \le c_2$ such that, for all $(\rho, T)$ with $\beta \left(C_1 \rho^{s_1/d} + C_2 \rho^{s_2/d}\right) > 1/2$ and $\rho > \rho_2$, we have
\[
    c_1 \rho^{2/d} \le \mu_{\rm free} (\rho, T) \le c_2 \rho^{2/d}.
\]

We now deduce similar estimates for $\mu$ and $k_*$. Using the pointwise estimate
\[
\frac{k^2}{2} - \mu - \| w  * g \|_{L^\infty} \le h(k) \le  \frac{k^2}{2} - \mu,
\]
together with Lemma~\ref{lem:estime_g*w}, we obtain that
\begin{equation} \label{eq:mu_and_k*}
2 \mu   \le k_*^2 \le 2 \mu+2 \left( C_1 \rho^{s_1/d} + C_2 \rho^{s_2/d} \right).
\end{equation}
Then, using again Lemma~\ref{lem:estime_g*w}, we have 
\[
    \frac{1}{\re^{\beta\left(k^2/2-\mu \right)}+1}\leq g \leq \frac{1}{\re^{\beta\left(k^2/2-C_1\rho^{s_1/d} - C_2 \rho^{s_2/d}-\mu \right)}+1}.
\]
After integration, and using the fact that $I_\beta(\cdot)$ is increasing, we obtain
\[
    \mu_{\rm free}(\rho, T)-\left( C_1 \rho^{s_1/d} + C_2 \rho^{s_2/d} \right)\leq \mu \leq \mu_{\rm free}(\rho, T).
\]
Since $\mu_{\rm free}$ behaves as $\rho^{2/d}$ to leading order, we deduce that $\mu$ also behaves as $\rho^{2/d}$ to leading order. Together with~\eqref{eq:mu_and_k*}, we deduce as wanted that $k_*$ behaves as $\rho^{1/d}$ to leading order.
\end{proof}

\subsubsection{Estimate in $\Omega_2$ in the critical case~\eqref{eq:bound_hat_w_critical}}
We now assume that $w(k) = \kappa| k |^{1-d}$ and $d\geq2$, in which case the integral in~\eqref{eq:integral_on_sphere} is no longer bounded. Following again ideas from~\cite{GonHaiLew-19}, we write $w(k) = w_a(k) + R_a(k)$ with
\[
    w_a(k) := \dfrac{1}{(k^2 + a^2)^{\frac{d-1}{2}}} \quad \text{and} \quad 
    R_a(k) := w(k) - w_a(k) = \dfrac{1}{| k |^{d-1}} - \dfrac{1}{(k^2 + a^2)^{\frac{d-1}{2}}},
\]
where $a> 0$ is a (small) parameter that we optimise at the end. This gives
\begin{equation} \label{eq:norm_Ag_s=1}
    \| A_g \| = \beta \| w*g(1-g) \|_{L^\infty} \le \beta \left\| w_a * g(1-g) \right\|_{L^\infty} + \beta \| R_a \ast g(1-g) \|_{L^\infty}.
\end{equation}
The last term of~\eqref{eq:norm_Ag_s=1} is controlled using the fact that $g(1-g) \le \frac14$, so that
\[
\beta \| R_a * g(1-g) \|_{L^\infty} \le \frac{\beta}{4} \int_{\R^d} R_a(k) \rd k
= \frac{\beta}{4} a \int_{\R^d} R_1(k) \rd k,
\]
where we used the fact that $R_a(k) = a^{d-1}R_1(ak)$ by scaling, and the fact that $R_1(k)$ goes as $k^{-(d+1)}$ at infinity and $|k|^{1-d}$ at $0$, hence is integrable over $\R^d$. 

For the first part of~\eqref{eq:norm_Ag_s=1}, we use again that 
$$g(1-g) \le \frac1{ 4 + \frac{\beta^2}{4} | k^2 - k_*^2 |^2},$$
so that
\begin{align}
 &\left\| w_a * g(1-g) \right\|_{L^\infty} \\
 & \qquad\qquad\le  \sup_{\ell \in \R^d} \left( \int_{0}^\infty \dfrac{ r^{d-1} \rd r}{4 + \frac{\beta^2}{4} | r^2 - k_*^2 |^2} 
 \int_{\SS^{d-1}} \frac{ \rd \omega }{ \left( |\ell-r \omega |^2 + a^2 \right)^{\frac{d-1}{2}}} \right) \nn \\
& \qquad\qquad\le  \int_{0}^\infty \dfrac{ r^{d-1} \rd r}{4 + \frac{\beta^2}{4} | r^2 - k_*^2 |^2} 
\sup_{\ell \in \R^d}  \left(  \int_{\SS^{d-1}} \frac{ \rd \omega }{ \left( |\ell-r \omega |^2 + a^2 \right)^{\frac{d-1}{2}}} \right) \nn\\
& \qquad\qquad=  \dfrac{1}{\beta k_*^{3}} \int_{0}^\infty \frac{\rd r}{ \frac{4}{\beta^2 k_*^4} +  \frac14 | r^2 - 1 |^2} 
\sup_{\ell' \in \R^d}  \left(  \int_{\SS^{d-1}} \frac{ \rd \omega }{ \left( |\ell'-\omega |^2 + \left(\frac{a}{k_*r}\right)^2 \right)^{\frac{d-1}{2}}} \right) \nn.
\end{align}

The last parenthesis is estimated with the following lemma, that we prove below.
\begin{lemma} \label{lem:estim_log}
We have
\begin{equation*}
 \sup_{\ell \in \R^d}  \left(  \int_{\SS^{d-1}} \frac{ \rd \omega }{ \left( |\ell-\omega |^2 + \lambda^2 \right)^{\frac{d-1}{2}}} \right)\leq C\left( 1 + | \log \lambda^{-1} | \right).
\end{equation*}
\end{lemma}
This gives
\begin{align*}
    \left\| w_a * g(1-g) \right\|_{L^\infty}  
    & \le \dfrac{1}{\beta^2 k_*^3} \int_{0}^\infty \dfrac{\rd r}{\frac{4}{\beta^2k_*^4} + \frac{1}{4} | r^2 - 1 |^2} \left( 1 + \left| \log \left( \frac{k_*r}{a} \right) \right| \right) \\
& \leq C \left(1+\left|\log \left( \dfrac{k_*}{a} \right) \right| \right)\left(\frac{1}{\beta^2 k_*^3} +\frac{1}{\beta k_*}\right).
    \end{align*}
Altogether, we have shown that for all $0 < a < 1$ and all $\beta$ and $k_*$, we have
\[
    \| A_g \| \le C\left(1+\left|\log \left( \dfrac{k_*}{a} \right) \right|\right)\left(\frac{1}{\beta k_*^3} +\frac{1}{k_*}\right)+Ca\beta.
\]
This leads us to choose $a = \alpha /\beta$ for a small enough constant $\alpha$. From Lemma~\ref{lem:estimate_k*} we know that $k_*$ is of the order of $\rho^{1/d}$. 
In the region $\Omega_2$ we take $\beta \rho^{1/d} >1/C_1$ and $\rho\geq \rho_c$. We deduce that $\| A_g \| <1$ whenever
\[
    \beta \le C \re^{ \alpha \rho^{1/d}},
\]
which concludes the proof in the case $s = 1$. It only remains to provide the

\begin{proof}[Proof of Lemma~\ref{lem:estim_log}]
When $| | \ell | - 1 | > \frac12$, the integrand is uniformly bounded for all $\lambda > 0$. We now assume that $1/2\leq |\ell|\leq 3/2$. In hyperspherical coordinates, our integral is proportional to
\begin{align*}
\int_{0}^{\pi/2} \dfrac{ \sin^{d-2} \theta \,\rd \theta }{ \Big(|\ell|^2-2|\ell|\cos(\theta)+1+\lambda^2 \Big)^{\frac{d-1}{2}} } 
& =
\int_{0}^{\pi/2} \dfrac{ \sin^{d-2} \theta \,\rd \theta }{ \Big(   |\ell - 1 |^2 + 4 |\ell| \sin^2 \frac{\theta}{2} +\lambda^2 \Big)^{\frac{d-1}{2}} } \\
& \le \int_{0}^{\pi/2} \dfrac{ \sin^{d-2} \theta \,\rd \theta }{ \Big( 2 \sin^2 \frac{\theta}{2} +\lambda^2 \Big)^{\frac{d-1}{2}} },
\end{align*}
where we used that $| \ell | \ge \frac12$ in the last inequality. Now, we use that for $0 \le \theta \le \frac{\pi}{2}$, we have $\sin \theta \le \theta$ and $\sin(\theta/2) \ge \frac{\sqrt{2}}{\pi}\theta$. We get the upper bound
\[
    \int_{0}^{\pi/2} \dfrac{ \theta^{d-2} \,\rd \theta }{ \Big( \frac{8}{\pi^2} \theta^2 +\lambda^2 \Big)^{\frac{d-1}{2}} }
     = \int_0^{\pi/(2\lambda)} \dfrac{ y^{d-2} \,\rd y }{ \Big( \frac{8}{\pi^2} y^2 +1 \Big)^{\frac{d-1}{2}} }.
\]
The integrand is bounded near $y = 0$, and behaves as $C y^{-1}$ to infinity, so the integral is bounded by $C(1 + | \log \lambda^{-1} |)$ as claimed.
\end{proof}

The proof of Theorem~\ref{th:uniqueness_positive_T} is now complete. \qed

\subsection{Proof of Theorem~\ref{th:uniqueness_paramagnetism_positive_T}, uniqueness for the spin-polarised problem}
We now turn to the proof of Theorem~\ref{th:uniqueness_paramagnetism_positive_T}. We write again $\widetilde{\Omega} = \widetilde{\Omega_1} \cup \widetilde{\Omega_2}$, with $\widetilde{\Omega_1} = \Omega_1$ as defined in~\eqref{eq:def:Omega1}, and 
\begin{equation*}
\widetilde{\Omega_2} := \left\{  (\rho, T) \in \R_+^* \times \R_+^*, \quad C_1 \rho^{s_1/d}+C_2 \rho^{s_2/d} > \frac{T}2, \quad T \re^{ \alpha \rho^{\frac{1}{d'}}} >  C \right\},
\end{equation*}
with $d' = d$ in the sub-critical case $1 < s < 2$, and $d' = 2d$ in the critical case $s_1=s_2 = 1$. We now use two different arguments in each region.

\subsubsection{Uniqueness in $\widetilde{\Omega_1}$} Let $(\rho, T) \in \widetilde{\Omega_1}$, so that the segment $[0, \rho] \times \{T\}$ is included in $\Omega$. Then, since the map $\rho' \mapsto E^\HF_\nospin(\rho', T)$ is strictly convex on $[0, \rho]$ (see Corollary~\ref{cor:branches}), we have, for all $t \in [0, 1/2]$,
\[
\frac12 E^\HF_\nospin(t \rho) + \frac12 E^\HF_\nospin((1 - t) \rho) \ge 
E^\HF_\nospin \left( \frac12 t \rho + \frac12 (1-t) \rho \right) = E^\HF_\nospin (\rho/2),
\]
and there is equality only for $t=1/2$. This implies that the minimum in~\eqref{eq:decoupled_energy} is attained only at $t=1/2$, hence that the minimiser of $E^{\rm HF}(\rho, T)$ is paramagnetic by Theorem~\ref{thm:spin_as_nospin}, with $g_{\rho/2, T}$ the unique minimiser of $E^{\rm HF}_\nospin(\rho/2, T)$.

\subsubsection{Uniqueness in $\widetilde{\Omega_2}$}  Let $(\rho, T) \in \widetilde{\Omega_2}$, and let
$g_0$ be any radial decreasing solution of~\eqref{eq:EulerLagrange_g_proof} with density $\rho/2$. We set $\gamma_0 = g_0 \bbI_2$, which has density $\rho$, and prove that $\gamma_0$ is the only minimiser of $\cE(\rho, T)$. 

Let $\gamma$ be any minimiser for $E^\HF(\rho,T)$. By Theorem~\ref{thm:spin_as_nospin}, we may take $\gamma$ diagonal without loss of generality, and write $\gamma = {\rm diag}(g_\spinup, g_\spindown)$ with $\rho_\spinup + \rho_\spindown = \rho_0$, where $\rho_{\uparrow,\downarrow} := (2 \pi)^{-d} \int_{\R^d} g_{\uparrow,\downarrow}$. We can compute the free energy difference as
\begin{multline*}
\cE^{\rm HF}(T,\gamma) -\cE^{\rm HF}(T,\gamma_0)\\
= \sum_{\sigma \in \{\spinup, \spindown\} }T\,\cH(g_\sigma,g_0)-\frac{1}{2(2\pi)^d}\iint_{\R^d\times\R^d}(g_\sigma-g_0)(k)(g_\sigma-g_0)(k')w(k-k')\,{\rm d}k\,{\rm d}k',
\end{multline*}
where
$$\cH(g,g_0)=\frac{1}{(2\pi)^d}\int_{\R^d}\left(g(k)\log\frac{g(k)}{g_0(k)}+(1-g(k))\log\frac{1-g(k)}{1-g_0(k)}\right)\,{\rm d}k$$
is the fermionic relative entropy. We use the inequality (see~\cite[Thm~1]{HaiLewSei-08}) 
$$ \forall 0 \leq x, y , \leq 1, \quad x\log\frac{x}{y}+(1-x)\log\frac{1-x}{1-y}\geq \frac{(x-y)^2}{2y-1}\log\frac{y}{1-y}.$$
Denoting $h_0(k) := k^2/2 - w*g_0 - \mu$, so that $g_0 = (1 + \re^{\beta h_0})^{-1}$, we find
\begin{equation*}
\cH(g,g_0) \geq \dfrac{1}{(2 \pi)^d} \int_{\R^d} \dfrac{\beta h_0}{ \tanh \left( \beta \frac{h_0}{2} \right)} (g-g_0)^2.
\end{equation*}
Using the simple bound $x \tanh(x/2)^{-1} \ge \max\{2, | x | \} \ge 1 + \frac12 | x |$ and the fact that $| h_0 | \ge  | k^2 - k_*^2|/2$ as we have seen in~\eqref{eq:estimate_with_laplacian}, we obtain
\[
\cH(g,g_0) \geq \dfrac{1}{(2 \pi)^d} \int_{\R^d} \left( 1 + \frac{1}{4} \beta | k^2 - k_*^2| \right) (g-g_0)^2 (k) \rd k.
\]
This leads to
\[
\cE^{\rm HF}(T,\gamma) -\cE^{\rm HF}(T,\gamma_0) \ge \sum_{\sigma \in \{ \spinup, \spindown \}} \frac12 \dfrac{1}{(2 \pi)^d} \left\langle (g_\sigma - g_0), \widetilde{\cL} (g_\sigma - g_0) \right\rangle,
\]
with
\[
\widetilde{\cL} g :=\left( 2T + \frac12 | k^2 - k_*^2|  \right) g - w*g. 
\]
We now prove that this operator is (strictly) positive, which leads to the equality $g_\sigma=g_0$. To this end we use an argument from our recent work~\cite{GonHaiLew-19} with Christian Hainzl. 

We first remark that it is enough to study the case where $w = \kappa | \cdot |^{s-d}$. Indeed, since $w\geq0$, the first eigenvector of $\widetilde\cL$ is necessarily positive. Therefore, by the variational principle, the pointwise bound $w(k)\leq\kappa_1|k|^{s_1-d} + \kappa_2 | k |^{s_2 - d}$ implies the bound on the bottom of the spectrum
$$\lambda_1(\widetilde\cL)\geq \lambda_1(\widetilde{\cL}_{s_1}) + \lambda_1(\widetilde{\cL}_{s_2})$$
with 
\[
    \widetilde{\cL}_{s} g := \left( T  + \frac14 | k^2 - k_*^2|  \right) g - \dfrac{\kappa}{| \cdot |^{d-s}} *g.
\]
Taking the Fourier transform, we see that $\widetilde{\cL}_{s} \ge 0$ if and only if
\[
     -\lambda_1\left(\frac14 | \Delta + k_*^2 | - \dfrac{\kappa}{c_{d,s} | x |^s}\right)\leq T.
\]
After scaling this is the same as
\begin{equation}  \label{eq:positivity_Hs}
     -\lambda_1\left( | \Delta + 1 | - \dfrac{\eps}{| x |^s}\right)\leq \frac{4T}{k_*^2},\qquad\text{with}\quad \eps=\frac{4\kappa}{c_{d,s}k_*^{2-s}}
\end{equation}
Note that from~\cite{LapSafWei-02,HaiSei-10,GonHaiLew-19} it is known that the operator on the left side always has a negative eigenvalue. We need an estimate on this eigenvalue. This is what has been accomplished in~\cite{HaiSei-10} for regular potentials and in~\cite{GonHaiLew-19} in the critical case $d=2,3$ and $s=1$. Following the exact same method, we can prove the 

\begin{lemma}[Estimate on the first eigenvalue of $|\Delta+1|-\eps|x|^{-s}$] \label{lem:estimateForHs}
    Let $1 \leq s < 2$ in dimension $d\geq2$. There is $C,\alpha > 0$ such that we have
    \[
         | \Delta - 1 | -  \dfrac{\eps}{| x |^s}   \ge - C \begin{cases}
\re^{-\tfrac{\alpha}\eps}&\text{for $1<s<2$,}\\
\re^{-\tfrac{\alpha}{\sqrt{\eps}}}&\text{for $s=1$},\\
                                                              \end{cases}
    \]
for all $0 < \eps\leq 1$.
\end{lemma}
    
\begin{proof}[Proof of Lemma~\ref{lem:estimateForHs}]
    The cases $d=2,3$ and $s=1$ have been handled in~\cite{GonHaiLew-19}. The proof is exactly the same in higher dimensions. The same argument indeed applies in the subcritical case $1<s<2$ and we only outline it here, following the notation in~\cite{GonHaiLew-19}. We have $| \Delta + 1 | - \eps | x |^{-s} \ge -E$ whenever
    \[
        I(E) := \dfrac{1}{(2 \pi)^d} \int_0^\infty \dfrac{r^{d-1} \cN(r) }{|r^2 -1 | + E} \rd r \le \frac{1}{\eps},
    \]
    where
    \[
        \cN(r) := \norm{\frac{1}{|x|^{\frac{s}{2}}} \int_{\SS^{d-1}}e^{\ri r\omega\cdot(x-y)}\,\rd\sigma(\omega) \frac{1}{|y|^{\frac{s}{2}}}}_{\rm op}= \dfrac{1}{r^{d-s}} \cN(1),
    \]
    by scaling. This gives
    \[
        I(E) = \dfrac{\cN(1)}{(2 \pi)^d} \int_0^\infty \dfrac{r^{s-1}}{|r^2 -1 | + E} \rd r.
    \]
    Since $1 < s < 2$, the integral is finite. Actually, when $E \to 0$, only the singularity at $r = 1$ diverges. Using computations similar to the ones used in Section~\ref{sssec:estimates_in_Omega2}, we deduce that, as $E \to 0^+$, we have $I(E) = \cN(1)(2 \pi)^{-d} \log(E^{-1}) (1 + o(1))$ and the result follows.
\end{proof}

Using Lemma~\ref{lem:estimateForHs} and Lemma~\ref{lem:estimate_k*} which states that $k_*\sim\rho^{1/d}$, this concludes the proof of Theorem~\ref{th:uniqueness_paramagnetism_positive_T}.
\qed

\begin{remark}[Numerical method for computing the phase diagram in Figure~\ref{fig:phaseDiagram}]\label{rmk:numerics}
We now briefly explain how we found numerically the minimiser of $E^\HF_\nospin$, which allowed us to plot the phase diagram in Figure~\ref{fig:phaseDiagram}. Our idea was to solve the Euler-Lagrange equation~\eqref{eq:EulerLagrange_mu} directly, and to look for all fixed points of $\cV_{\mu}$ in~\eqref{eq:Hammerstein_map}, for all Fermi levels $\mu$. Actually, since the potentials $V$ are slowly decreasing, we choose to work with the functions $g$, and look for fixed points of
\[
    g = \cG_{\mu, T}(g), \quad \text{with} \quad \cG_{\mu, T}(f) := \dfrac{1}{\re^{\beta (k^2/2 -\mu - g*w)} + 1}.
\]
As before, we can define
\[
    g_{\rm min}[T, \mu]:= \lim_{n \to \infty} \cG_{\mu}^{(n)}(\bnull), \quad \text{and} \quad
    g_{\rm max}[T, \mu]:= \lim_{n \to \infty} \cG_{\mu}^{(n)}(\bone),
\]
which correspond respectively to the minimal and the maximal fixed points of $\cG_{\mu}$. Both $g_{\rm min}$ and $g_{\rm max}$ can be computed efficiently by iterating the map $\cG_{\mu}$. In the case where $\cG_{\mu}$ has a unique fixed point, then $g_{\rm min} = g_{\rm max}$, and we are done. If $g_{\rm min} \neq g_{\rm max}$, then we expect a middle fixed point of $\cG_{\mu}$ (see Figure~\ref{fig:rho_mu}). To find this middle fixed point, we used a string method: we construct a continuous initial path $g^0(t)$ with $g^0(t = 0)= g_{\rm min}$ and $g^0(t = 1) = g_{\rm max}$, and we define iteratively the path $g^{n+1}(t):= \cG_{\mu}[ g^{n}(t)]$.
After some iterations, the whole path has converged to some $g^\infty(t)$, and we look for the middle fixed point of $\cG_{\mu}$ on this path.
In practice, the path is re-parametrised at each iteration, and sampled uniformly in order to avoid the points from falling into the two  valleys~\cite{CanGalLew-06}. 
\end{remark}

\appendix
\section{Proof of Lemma~\ref{lem:well-posed}}\label{sec:proof_lemma_well_posed}
The energy $\cE^{\rm HF}_T$ is well defined and bounded from below on the set of all the fermionic density matrices $\gamma$ with $\rho_\gamma=\rho<\ii$. Indeed, for $w=W_1+W_\ii\in L^1(\R^d)+L^\ii(\R^d)$, we can estimate
\begin{multline*}
\iint_{\R^d\times\R^d} |w(\bk - \bk')| \tr_{\C^2} \left[ \gamma (\bk) \gamma(\bk') \right] \rd \bk\, \rd \bk'\\
\leq (2\pi)^{2d}\norm{W_\ii}_{L^\ii(\R^d)}\rho_\gamma^2+(2\pi)^{d}\norm{W_1}_{L^1(\R^d)}\rho_\gamma, 
\end{multline*}
since 
$$\tr_{\C^2}[ \gamma (\bk) \gamma(\bk') ]\leq \min\Big\{\tr_{\C^2}[ \gamma (\bk)]\;,\;\tr_{\C^2}[ \gamma (\bk)]\, \tr_{\C^2}[\gamma(\bk')]\Big\}.$$
At $T>0$ we control the entropy by using the Fermi-Dirac non-interacting solution for half the kinetic energy and we deduce that 
\begin{multline}
\frac{1}{2} \int_{\R^d} k^2 \tr_{\mathbb{C}^2}\gamma(\bk)\, \rd \bk +T \int_{\R^d} \tr_{\mathbb{C}^2}S(\gamma(\bk)) \, \rd \bk\\
\geq \frac{1}{4} \int_{\R^d} k^2 \tr_{\mathbb{C}^2}\gamma(\bk)\, \rd \bk-2T\int_{\R^d}\log\left(1+e^{-k^2/(4T)}\right)\,\rd\bk.\label{eq:coercive}
\end{multline}
In all cases we have proved that 
$$\cE^{\rm HF}_T(\gamma)\geq \frac{1}{4} \int_{\R^d} k^2 \tr_{\mathbb{C}^2}\gamma(\bk)\, \rd \bk-C$$
for a constant $C$ depending on $T$ and $\rho_\gamma=\rho$, which stays bounded in the limit $T\to0^+$.
Let then $\{\gamma_n\}_{n\geq1}$ be a minimising sequence for the problem~\eqref{eq:GS_HF_free_energy}, that is, such that $\cE^{\rm HF}(\gamma_n, T)\to E^{\rm HF}(\rho,T)$. This sequence is bounded in $L^1(\R^d)\cap L^\ii(\R^d)$, hence converges weakly--$\ast$ in that space to a fermionic state $\gamma$, after extraction of a subsequence. The bound~\eqref{eq:coercive} implies that the sequence is tight in $L^1(\R^d)$ and hence the convergence must be strong in $L^p(\R^d)$ for all $1\leq p<\ii$, by interpolation. In particular, we get that $\rho_\gamma=\rho$. This is sufficient to pass to the limit in the exchange term. The kinetic energy being lower semi-continuous, we conclude that $\cE^{\rm HF}(\gamma, T)= E^{\rm HF}(\rho,T)$ and hence $\gamma$ is a minimiser. That any minimiser solves the mentioned nonlinear equation is standard and the arguments are all the same for the no-spin problem. 

Now, if in addition $w$ is radial non-increasing, we can prove that any minimiser $g$ for the no-spin problem~\eqref{eq:GS_HF_free_energy_no_spin} is also radial non-increasing. Let $g^*$ denote the symmetric rearrangement of a function $g$~\cite[Chap.~3]{LieLos-01}. Then $g^*$ has the same average density and the same entropy as $g$, by~\cite[Eq.~(3) \& (4)]{LieLos-01}. Also,
$$\int_{\R^d}k^2g^*(\bk)\,\rd\bk\leq \int_{\R^d}k^2g(\bk)\,\rd\bk$$
with equality if and only if $g$ is already radial non-increasing by~\cite[Thm.~3.4]{LieLos-01} and
$$\iint_{\R^d\times\R^d} w(\bk - \bk')  g(\bk)g(\bk') \rd \bk\, \rd \bk'\leq \iint_{\R^d\times\R^d} w(\bk - \bk')  g^*(\bk)g^*(\bk') \rd \bk\, \rd \bk'$$
by the Riesz rearrangement inequality~\cite[Thm.~3.7]{LieLos-01}. From this we conclude that the minimisers of the no-spin problem $E_\nospin^{\rm HF}(\rho,T)$ ought to be radial non-increasing. 
This concludes the proof of Lemma~\ref{lem:well-posed}.\qed

\section{All the critical points are radial decreasing}\label{app:moving_planes}

In this section, we prove that all the fixed points of $\cV_{\mu, T}$ are positive radial decreasing, using the moving plane method~\cite{GidNiNir-81}. We assume that $V$ is regular enough.

\begin{lemma}[All the critical points are radial-decreasing]
    Let $0\neq w\in L^1(\R^d)+L^\infty(\R^d)$ be a non-negative radial non-increasing function. Let $\beta > 0$ and $\mu \in \R$. Then any solution $V \in W^{1,\ii}(\R^d)$ to the equation
    \begin{equation} \label{eq:fixed_point_appendix}
        V = w* \left(  \dfrac{1}{\re^{ \beta (k^2/2 - \mu - V)} + 1}  \right)
    \end{equation}
    is positive radial decreasing.
\end{lemma}

We assume here that $V$ is in $W^{1,\ii}(\R^d)$. This regularity can easily be proved for particular choices of $w$, including for instance $w(k) = | k |^{s - d}$ with $1 \le s < d$ or a finite sum of such functions. 

\begin{proof}
We adapt here the moving plane method to our case, and give a simple self-contained proof. Let $V \in W^{1,\infty}(\R^d)$ be a solution to~\eqref{eq:fixed_point_appendix}. By construction $V$ is always positive. We set
\[
    g(k) := \dfrac{1}{\re^{ \beta (k^2/2 - \mu - V(k))} + 1}, \quad \text{so that} \quad
    V = w*g.
\]
Since $V \in L^\infty(\R^d)$, we have $g \in L^1(\R^d) \cap L^\infty(\R^d)$. The second equation then shows that $V$ (hence $g$) is in fact continuous. Differentiating the equation we find that $\nabla V$ and $\nabla g$ are continuous and tend to zero at infinity.

Let us assume by contradiction that $V$ and $g$ are not radial. Then there is a plane $\Sigma_0$ going through the origin such that the reflection of $g$ over $\Sigma_0$ differs from $g$. Without loss of generality, we may assume that $\Sigma_0 = \{ k_1 = 0\}$, and we set
 $g_0(k_1, k_\perp) := g(-k_1, k_\perp)$, which solves the same fixed point equation as $g$. The function $g_0 - g$ is a non-zero odd function, hence there is $k^* \in \R^d$ with $g_0(k^*) < g(k^*)$. Since $g_0(0, k_\perp) = g(0, k_\perp)$, one has $k_1^* \neq 0$, and, without loss of generality, we may assume $k_1^* > 0$ (otherwise, we replace $g$ by $g_0$).

For $\lambda \ge 0$, we set 
\[
    \Sigma_\lambda := \{k \in \R^d, \ k_1 = \lambda \}, \quad \text{and} \quad \Sigma_\lambda^+ := \{k \in \R^d, \ k_1 > \lambda\}.
\]
We denote by $g_\lambda(k_1, k_\perp) := g(2 \lambda - k_1, k_\perp)$ the reflection of $g$ over the plane $\Sigma_\lambda$, and by $V_\lambda = w*g_\lambda$ the associated potential. We claim that for all $\lambda > 0$, we have $g_\lambda \ge g$ on $\Sigma_\lambda^+$. At the limit $\lambda \to 0$, this contradicts the fact that $g_0(k^*) < g(k^*)$ with $k^* \in \Sigma_0^+$.
First, let us prove that there is $\lambda_C > 0$ such that for all $\lambda \ge \lambda_C$, we have $g_\lambda \ge g$ on $\Sigma_\lambda^+$.
We have
\[
    g_\lambda(k) = \left( \re^{\beta ( (2 \lambda - k_1)^2 + k_\perp^2 - \mu - V_\lambda)} +1   \right)^{-1}
        =  \left( \re^{\beta ( 4 \lambda(\lambda - k_1) + k^2 - \mu - V_\lambda)} +1   \right)^{-1}.
\]
This gives
\begin{equation} \label{eq:diff_glambda-g}
    (g_\lambda - g)(k) = (1 - g(k)) g_\lambda(k) \left( 1 - \re^{-\beta (k_1-\lambda) \left[ 4 \lambda + \frac{V_\lambda(k) - V(k)}{k_1-\lambda}   \right]} \right).
\end{equation}
Since $0 < g < 1$ and $0 < g_\lambda < 1$, we deduce that $g_\lambda - g \ge 0$ on $\Sigma_\lambda^+$ if and only if 
\begin{equation*}
    \forall k \in \Sigma_\lambda^+, \quad \dfrac{V_\lambda(k) - V(k)}{k_1-\lambda} + 4 \lambda\geq0.
\end{equation*}
After the change of variable $(k_1, k_\perp) \to (k_1 - \lambda, k_\perp)$, this is also equivalent to
\begin{equation} \label{eq:equivalence_glambda-g}
    \forall k \in \Sigma_0^+, \quad U(\lambda, k) := \dfrac{V(\lambda - k_1, k_\perp) - V(\lambda + k_1, k_\perp)}{k_1} \ge - 4 \lambda.
\end{equation}
Since $\nabla V$ is bounded, $U(\cdot, \cdot)$ is uniformly bounded by some constant $C > 0$.
Setting $\lambda_C = C+1$, we get that for all $\lambda> \lambda_C$, and all $k_1 > \lambda$, we have $(g_\lambda - g) (k) \ge 0$, as claimed.

We now set 
\[
    \lambda^* := \inf \left\{ \lambda > 0, \  g_\lambda \ge g \ \text{on} \ \Sigma_\lambda^+   \right\}.
\]
By continuity, we have $g_{\lambda^*} \ge g$ on $\Sigma_{\lambda^*}^+$. In particular, since $g_0(k^*) < g(k^*)$, one must have $\lambda^* > 0$. Let us prove that $V_{\lambda^*} \geq V$ on $\Sigma_{\lambda^*}^+$. Indeed, we have, with a change of variable (we set $\lambda = \lambda e_1$)
\[
    V(k) = \int_{\R^d} w(k - \ell) g(\ell) \rd \ell 
    = \int_{\Sigma_\lambda^+} w(k - \ell) g(\ell) \rd \ell + \int_{\Sigma_\lambda^+} w(k + \ell - 2 \lambda) g_\lambda(\ell) \rd \ell.
\]
This gives, for all $\lambda \ge 0$,
\begin{equation} \label{eq:diff_Vlambda-V}
    (V_\lambda - V)(k) = \int_{\Sigma_\lambda^+} \left[w(k - \ell) - w(k + \ell - 2 \lambda) \right] \left( g_\lambda - g \right) \rd \ell.
\end{equation}
For $k_1 > \lambda$ and $\ell_1 > \lambda$, we have
\[
    | k_1 - \ell_1 | < | k_1 - \lambda | + | \lambda_1 - \ell | = k_1 + \ell_1 - 2 \lambda, \quad \text{hence} \quad
    |k - \ell | \le | k + \ell - 2 \lambda|.
\]
Since $w$ is radial decreasing, it implies that the function in the brackets appearing in~\eqref{eq:diff_Vlambda-V} is positive on $\Sigma_\lambda^+ \times \Sigma_\lambda^+$. Hence if $g_\lambda \ge g$ on $\Sigma_\lambda^+$, we deduce that $V_\lambda \geq V$ on $\Sigma_\lambda^+$. Applying this at $\lambda = \lambda^*$, this proves that $V_{\lambda^*} \geq V$ on $\Sigma_{\lambda^*}^+$.

In particular, we have $U(\lambda^*, k) \geq 0$ for all $k \in \Sigma_0^+$, where $U$ was defined in~\eqref{eq:equivalence_glambda-g}. By the uniform continuity of $U$, there is $\lambda' < \lambda^*$ such that $U(\lambda, k) > -4\lambda$ for all $\lambda' < \lambda < \lambda^*$ and all $k \in \Sigma_0^+$. 
This proves that~\eqref{eq:equivalence_glambda-g} is satisfied for all $\lambda > \lambda'$, which contradicts the definition of $\lambda^*$.

We have proved that both $V$ and $g$ are radial. To show that they are radial decreasing, we repeat the argument, and obtain that
\[
    \forall \lambda > 0, \ \forall k \in \Sigma_\lambda^+, \quad V_\lambda(k) \geq V(k).
\]
Let $0 < r < R$. We take $\lambda = \frac12 (R+ r) > 0$ and $k = (R,0) \in \Sigma_\lambda^+$, and get that $V(r) \geq V(R)$. Then, due to the strict monotonicity of $k^2$, $g$ is decreasing. Thus, $V$ is also decreasing. 
\end{proof}


\newcommand{\etalchar}[1]{$^{#1}$}

\end{document}